\pdfoutput=1
\documentclass[
pdflatex,sn-mathphys-num]{sn-jnl}

\usepackage{graphicx}
\usepackage{xcolor}
\usepackage{amsmath,amssymb,amsfonts}
\usepackage{amsthm}
\usepackage[title]{appendix}
\usepackage{braket}
\usepackage{dcolumn}
\usepackage{bm}
\usepackage{bookmark}
\hypersetup{hidelinks}
\usepackage{booktabs}

\theoremstyle{thmstyleone}
\newtheorem{theorem}{Theorem}[section]
\begin{document}

\title[CFTs with Large Virasoro Primary Gaps]{CFTs with Large Virasoro Primary Gaps}

\author*[2]{\fnm{Ziming} \sur{Ji}}\email{zi.ji@northeastern.edu}
\author*[1]{\fnm{Jiahua} \sur{Tian}}\email{jtian1905@gmail.com}

\affil[1]{\orgdiv{School of Physics}, \orgname{East China Normal University}, \orgaddress{\city{Shanghai}, \postcode{200241}, \country{China}}}
\affil[2]{\orgdiv{Department of Physics}, \orgname{Northeastern University}, \orgaddress{\city{Boston}, \state{MA}, \postcode{02115}, \country{USA}}}

\abstract{We construct the first explicit examples of compact unitary bosonic full conformal field theories (CFTs) in two dimensions with a unique vacuum and an all-spin Virasoro-primary gap greater than two. Our constructions use asymmetric orbifolds of lattice CFTs built from even unimodular lattices of sufficiently large minimum norm, ensuring that no charged state lies below the desired gap. The gauged symmetries combine a common orthogonal action on the left- and right-moving currents with independent sign reversals; the latter project out the dimension-two scalar primary $\sum_i J_L^iJ_R^i$ that survives when the symmetry acts identically on both chiral halves. We prove that the gauging is anomaly-free and determine the exact gaps by analyzing the invariant untwisted spectrum and all twisted sectors. From lattices constructed using ternary quadratic-residue codes, we obtain an infinite family with gap three at $(c_L,c_R)=(p+1,p+1)$ for primes $p\ge71$ with $p\equiv23\pmod{24}$. From Barnes--Wall lattices, we obtain an infinite family with gap four at $(c_L,c_R)=(2^m,2^m)$ for odd $m\ge7$. We also obtain two additional examples, with gap three at $(c_L,c_R)=(48,48)$ and gap four at $(c_L,c_R)=(512,512)$. }

\maketitle

\tableofcontents

\section{\label{sec:intro}Introduction}

The search for two-dimensional conformal field theories (CFTs) with large Virasoro primary gaps connects several fundamental questions in physics. This problem has mathematical roots in the study of extremal vertex operator algebras~\cite{Hoehn1995} and is motivated by determining which spectral gaps allowed by modular-bootstrap bounds are realizable~\cite{Hellerman2011,CollierLinYin2018}, finding renormalization-group fixed points without relevant or marginal scalar deformations~\cite{DMS1997,Nakayama2015}, and understanding the spectral requirements for holographic descriptions of three-dimensional gravity~\cite{Witten2007,HKS2014}. The existence of a compact unitary bosonic full CFT with a unique vacuum and a Virasoro primary gap greater than two was posed as an open problem in~\cite{KRR2026}. In this work, we resolve this existence problem by constructing explicit infinite families of such theories as nonfactorizing lattice orbifolds, with exact primary gaps three and four at arbitrarily large central charges. In~\cite{KRR2026}, a holomorphic CFT with gap four at $c=128$ was obtained from the orbifold of the rank-$128$ Barnes--Wall lattice VOA by the extraspecial group $2_+^{1+14}$, conditional on the conjectured vanishing of its chiral anomaly; our Barnes--Wall family instead gives nonfactorizing full CFTs with gap four at $c=2^m$ for every odd $m\ge7$, including $c=128$, without relying on this conjecture, because under correlated gauging the two chiral anomalies cancel without either having to vanish.

Central to our constructions is the combination of lattice geometry with carefully chosen finite symmetries. We begin with lattices with large minimum norm, using the standard relation between lattice-vector lengths and conformal weights to exclude light charged states~\cite{KRR2026}. Low-dimension primary fields can nevertheless arise from oscillator excitations, including currents and their composites. To remove these fields, we gauge a finite symmetry group $G$ of the lattice CFT, retaining invariant states and introducing sectors with symmetry-twisted boundary conditions~\cite{DHVW1985,DHVW1986}. In a full lattice CFT, identical orthogonal actions on the left- and right-moving currents leave an invariant scalar primary of dimension two~\cite{KRR2026}. We overcome this obstruction by adjoining independent sign reversals of the left- and right-moving currents, within the framework of asymmetric orbifolds~\cite{NSV1987}. The essential challenge is to eliminate the unwanted primaries while ensuring that the twisted sectors introduce no states below the desired gap and that the symmetry can be gauged consistently. The latter requires cancellation of the symmetry anomaly, which is the quantum obstruction to gauging~\cite{Bhardwaj:2017xup}.

We determine the low-dimension invariant spectrum using finite-group character theory and invariant tensors~\cite{GK2019,GK2021,KRR2026}, identifying which surviving states are Virasoro primaries. We verify level matching conditions in the twisted sectors~\cite{Vafa1986,vEMS2020,KRR2026,LinShao2021} and use group cohomology to rule out any remaining obstruction to gauging the full symmetry group~\cite{Bhardwaj:2017xup,Brown1982,JFT2020}.

Our constructions are carried out in the standard framework of finite gauging of two-dimensional CFTs~\cite{Bhardwaj:2017xup,FFRS2009,Runkel:2022fzi}. As is customary, we assume that the parent lattice theories, together with their invertible symmetry defects, define correlators on arbitrary surfaces compatible with sewing and unitarity. In this framework the anomaly is the class of the defect associator and trivializing it permits consistent choices of defect junctions, and the gauged correlators are defined by inserting and summing defect networks.

Our four constructions, namely two infinite families and two additional examples, use the same type of asymmetric action, in which the gauged symmetry acts differently on the left- and right-moving currents.
A finite group $K$ acts faithfully on an even unimodular lattice $L$ of rank $D$ through $r:K\to O(L)$, and a central involution $z\in K$ acts as $r(z)=-I_D$. With $A=K/\langle z\rangle$, we gauge the fiber product
\begin{equation}
    \widetilde G=K\times_AK
    =\{(k,z^\varepsilon k)\mid k\in K,\ \varepsilon\in\{0,1\}\},
    \label{eq:intro-fiber-product}
\end{equation}
whose element $(k_L,k_R)$ rotates the left- and right-moving currents by $r(k_L)$ and $r(k_R)$, respectively. Thus the two chiral rotations agree up to sign, while the central elements $(z,1)$ and $(1,z)$ reverse the left- and right-moving currents independently.
As an abstract group, $\widetilde G\cong K\times\mathbb Z_2$ through $(k,z^\varepsilon k)\mapsto(k,\varepsilon)$ (see~\eqref{eq:QR-group} and~\eqref{eq:GtildeAction}).

The parent theory is either the Narain CFT $\mathcal N(L)$, whose momentum and winding both take values in $L$, or the split theory $V_L\otimes\overline{V_L}$, whose left and right charges take values in $L$ independently. For $\mathcal N(L)$ the element $(k,z^\varepsilon k)$ acts by the lattice isometry $r(k)$ on both momentum and winding, combined with $t^\varepsilon$, where $t$ exchanges momentum and winding and therefore reverses the right-moving currents. We summarize our main result as follows.

\begin{theorem}[Main result]
\label{thm:main}
Let $p\ge71$ be a prime with $p\equiv23\pmod{24}$ and let $m\ge7$ be odd. Consider the following parents $\mathcal C$, lattices $L$ of rank $D$ and groups $K$.
\begin{enumerate}
    \item $\mathcal C=\mathcal N(L)$, where $L=L^{\mathrm{QR}}_{p+1}$ is the even unimodular lattice of rank $p+1$ built from the extended ternary quadratic-residue code of length $p+1$, and $K=\mathrm{SL}_2(p)$ acts by signed permutations of the coordinates labelled by $\mathbb P^1(\mathbb F_p)$ (see~\eqref{eq:qr-lattice} and~\eqref{eq:QR-signed-action}).
    \item $\mathcal C=\mathcal N(L)$, where $L=P_{48q}$ is the extremal even unimodular lattice of rank $48$ with automorphism group $K=\operatorname{Aut}(P_{48q})\cong\mathrm{SL}_2(47)$ (see~\eqref{eq:P48q-data} and~\eqref{eq:P48q-current-action}).
    \item $\mathcal C=\mathcal N(L)$, where $L=E_8^{\otimes3}$ is the tensor cube of $E_8$, of rank $512$, and $K=\bigl(W(E_8)'\bigr)^{\times3}/\mathbb Z_2^2$ acts by $[h_1,h_2,h_3]\mapsto h_1\otimes h_2\otimes h_3$ (see~\eqref{eq:T512-lattice} and~\eqref{eq:T512-current-action}).
    \item $\mathcal C=V_L\otimes\overline{V_L}$, where $L=BW(m)$ is the Barnes--Wall lattice of rank $2^m$ and $K\cong2_+^{1+2m}$ is the extraspecial subgroup of $\operatorname{Aut}(BW(m))$ (see~\eqref{eq:BW-fiber-product} and~\eqref{eq:SES_NGA}).
\end{enumerate}
In each case the fiber product $\widetilde G$ in~\eqref{eq:intro-fiber-product} lifts to a faithful unitary action on $\mathcal C$ with vanishing anomaly (see Theorems~\ref{thm:QR-lift}, \ref{thm:Narain-anomaly} and~\ref{thm:BW-anomaly}). The orbifold $\mathcal C/\widetilde G$ is a compact unitary bosonic nonfactorizing full CFT with a unique vacuum and $(c_L,c_R)=(D,D)$, and its Virasoro primary gap is three in cases~1 and~2 (see Theorems~\ref{thm:QR-gap} and~\ref{thm:P48q-gap}), and four in cases~3 and~4 (see Theorems~\ref{thm:T512-gap} and~\ref{thm:BW-gap}).
\end{theorem}

The four constructions are summarized in Table~\ref{tab:results}.

\begin{table}[htbp]
\centering
\caption{Summary of the orbifolds $\mathcal C/\widetilde G$ in
Theorem~\ref{thm:main}. Here $c=c_L=c_R$, and
$\Delta_{\mathrm{gap}}$ is the exact all-spin Virasoro-primary gap
of the orbifold.}
\label{tab:results}
\renewcommand{\arraystretch}{1.15}
\setlength{\tabcolsep}{5pt}
\begin{tabular*}{\textwidth}{@{\extracolsep{\fill}}lllcc@{}}
\toprule
Lattice $L$ & Parent $\mathcal C$ & Group $K$
    & $c$ & $\Delta_{\mathrm{gap}}$ \\
\midrule
$L^{\mathrm{QR}}_{p+1}$
    & $\mathcal N(L)$
    & $\mathrm{SL}_2(p)$
    & $p+1$ & $3$ \\
$P_{48q}$
    & $\mathcal N(L)$
    & $\mathrm{SL}_2(47)$
    & $48$ & $3$ \\
\addlinespace[2pt]
$E_8^{\otimes3}$
    & $\mathcal N(L)$
    & $\bigl(W(E_8)'\bigr)^{\times3}/\mathbb Z_2^2$
    & $512$ & $4$ \\
$BW(m)$
    & $V_L\otimes\overline{V_L}$
    & $2_+^{1+2m}$
    & $2^m$ & $4$ \\
\botrule
\end{tabular*}
\footnotetext{In every row,
$\widetilde G=K\times_AK\cong K\times\mathbb Z_2$, with
$A=K/\langle z\rangle$.
In the first row, $p\ge71$ is prime with $p\equiv23\pmod{24}$;
in the last row, $m\ge7$ is odd.}
\end{table}

The fiber product~\eqref{eq:intro-fiber-product} controls the neutral untwisted states, which carry no lattice charge and consist entirely of oscillator excitations. The central elements $(z,1)$ and $(1,z)$ act on them by $(-1)^{n_L}$ and $(-1)^{n_R}$, where $n_L$ and $n_R$ count the left- and right-moving oscillators, so every invariant neutral state has an even number of oscillators on each side. Hence no nonzero combination of the currents $J_L^i$, $J_R^i$ or of the dimension-two states $J_L^iJ_R^j$ is invariant. In particular, the gauging removes the scalar $\sum_iJ_L^iJ_R^i$, which survives identical actions on the two sides. Since $r$ is irreducible over $\mathbb R$, the only symmetric matrices commuting with $r(K)$ are multiples of the identity, so the only invariant neutral states of dimension two are the two stress tensors.
At weights $(3,0)$ and $(0,3)$, the invariant neutral primaries correspond to the antisymmetric matrices commuting with $r(K)$. In cases~1 and~2 the Paley matrix of the quadratic residues spans these matrices, whereas $r$ is absolutely irreducible in cases~3 and~4 and there are none.

For the twisted sectors of the elements $(k,zk)$, which rotate the two sides with opposite signs, we pair the eigenvalues of $r(k)$ and $-r(k)$ and obtain $\Delta\ge D/16$ in every case. In case~2 this bound is attained, and the twisted sectors of $(1,z)$ and $(z,1)$ contribute further primaries of dimension three. For the diagonal elements $(k,k)$ with $k\ne1$, we instead derive a bound specific to each representation.
Combining these bounds with $\Delta\ge\min(L)/2$ for the charged untwisted states, we obtain the gaps in Theorem~\ref{thm:main}.

Although the independent signs remove every state $J_L^iJ_R^j$, the
contraction of the traceless parts of $J_L^iJ_L^j$ and $J_R^iJ_R^j$
remains invariant, because the two chiral rotations agree up to sign.
This contraction is a nonzero primary of weights $(2,2)$, and we use it
to prove that the resulting CFTs are nonfactorizing.

The four constructions differ in their charges and in how the anomaly cancels. The Narain parents glue the left- and right-moving charges through momentum and winding, whereas the split parent has independent left and right charges. For the Narain parents we show that the anomaly vanishes on the geometric subgroup $\{(k,k)\}\cong K$ and on $\langle t\rangle$, and that the possible mixed anomaly between them also vanishes, whereas for the split parent the two chiral anomalies are inflated from the same class on $A$ and cancel.

The work is organized as follows. In section~\ref{sec:conventions} we fix conventions and review the necessary preliminaries. In section~\ref{sec:CFTs} we construct the quadratic-residue family of full CFTs with gap three, beginning at $(c_L,c_R)=(72,72)$. In section~\ref{sec:P48q} we extend the same orbifold argument to the lattice $P_{48q}$ at $(c_L,c_R)=(48,48)$ and explain the differences from the quadratic-residue family. In section~\ref{sec:T512} we apply the same construction to the tensor lattice $E_8^{\otimes3}$ and obtain a full CFT with gap four at $(c_L,c_R)=(512,512)$. In section~\ref{sec:full-BW} we construct an infinite Barnes--Wall family of full CFTs with gap four. Finally, in section~\ref{sec:conclusion} we summarize our results and discuss open questions. The proof of anomaly cancellation for the Narain constructions of sections~\ref{sec:CFTs}--\ref{sec:T512} is given in Appendix~\ref{app:Narain-anomaly}.

\section{\label{sec:conventions}Conventions and Preliminaries}

In this section we fix conventions and review the relevant preliminaries. We use vertex operator algebra (VOA) terminology for the chiral halves of the lattice CFTs. In particular, the chiral halves of the split constructions are holomorphic lattice VOAs. For the full theories below, we establish unitarity by anomaly-free finite gauging of Narain CFTs by symmetries that are compatible with the positive inner product~\cite{Narain1986,NarainSarmadiWitten1987,NSV1987,DLWW2024}.

We write $\mathbf 1$ for the vacuum and $L_n,\bar L_n$ for the two Virasoro algebras of a full CFT $\mathcal C$ with state space $\mathcal H$. The simultaneous $L_0,\bar L_0$ eigenspace is $\mathcal H_{h,\bar h}$, with scaling dimension $\Delta=h+\bar h$ and spin $s=h-\bar h$~\cite{DMS1997}. All full theories considered here are compact and bosonic, with a unique vacuum and $c_L=c_R=c$. Thus $c$ denotes either chiral central charge. Compactness means that the spectrum is discrete, with only finitely many states below any fixed scaling dimension~\cite{KaidiPerlmutter2020}. We define
\begin{align}
\mathcal P_{h,\bar h}(\mathcal H)
&=\{v\in\mathcal H_{h,\bar h}\mid L_nv=\bar L_nv=0\text{ for all }n>0\},\notag\\
\Delta_{\mathrm{gap}}(\mathcal C)
&=\min\{h+\bar h>0\mid\mathcal P_{h,\bar h}(\mathcal H)\ne0\}.
\label{eq:full-gap-definition}
\end{align}
This definition counts primaries of every allowed spin, including purely chiral primaries. Local fields in the bosonic full CFTs considered here have integer spin, $s=h-\bar h\in\mathbb Z$. However, the unprojected twisted sectors introduced below need not satisfy this condition~\cite{LinShao2021}. The positive-weight condition in~\eqref{eq:full-gap-definition} excludes the vacuum vector. At the central charges considered here, no nonzero positive-weight Virasoro vacuum descendant is primary. We establish $\Delta_{\mathrm{gap}}\ge d$, with $d=3$ or $4$, by showing that all states below total dimension $d$ belong to the Virasoro vacuum representation. We then identify a primary of total dimension $d$ to establish equality.

We normalize characters and partition functions by
\begin{align}
\operatorname{ch}_V(\tau)&=\operatorname{Tr}_V q^{L_0-c/24},\notag\\
Z_{\mathcal C}(\tau,\bar\tau)&=\operatorname{Tr}_{\mathcal H}
q^{L_0-c_L/24}\bar q^{\bar L_0-c_R/24},
\label{eq:character-convention}
\end{align}
where $q=e^{2\pi i\tau}$, $\bar q=e^{-2\pi i\bar\tau}$, and $\operatorname{Im}\tau>0$~\cite{DMS1997}. A graded dimension without the vacuum prefactor is written explicitly as $\operatorname{Tr}_V q^{L_0}$. We use $\eta(\tau)=q^{1/24}\prod_{n\ge1}(1-q^n)$ for the Dedekind eta function.

For a positive-definite even lattice $L$ of rank $D$, we write $\min(L)=\min_{0\ne\lambda\in L}(\lambda,\lambda)$ for its minimum squared norm. Its lattice VOA has the standard realization
\begin{equation}
V_L=M(1)\otimes\mathbb C_\epsilon[L],\qquad
c(V_L)=D,\qquad h(e^\lambda)=\frac{(\lambda,\lambda)}2.
\label{eq:lattice-voa-convention}
\end{equation}
Here $M(1)$ is the Heisenberg vertex operator algebra and $\mathbb C_\epsilon[L]$ is the group algebra associated to $L$ twisted by the 2-cocycle $\epsilon$~\cite{FLM1988,BK2004}. We use an orthonormal basis $a^i$ of $L\otimes_{\mathbb Z}\mathbb C$ with
\begin{equation}
[a_m^i,a_n^j]=m\delta^{ij}\delta_{m+n,0},\qquad
\omega=\frac12\sum_{i=1}^{D}(a_{-1}^i)^2\mathbf 1.
\label{eq:heisenberg-convention}
\end{equation}
Nonzero lattice charges in $V_L$ therefore have weight at least $\min(L)/2$. If $L = L^*$, then $V_L$ is holomorphic~\cite{FLM1988,BK2004}. Later we will use both tensor products and orthogonal direct sums of lattices. For the former we have $L\otimes L'$ with bilinear form $(x\otimes y,x'\otimes y')=(x,x')(y,y')$ and the ranks multiply, whereas for the latter the corresponding VOAs tensor as $V_{L\oplus L'} = V_L \otimes V_{L'}$.

A Narain CFT is specified by an even unimodular momentum lattice $\Gamma\subset\mathbb R^D\oplus\mathbb R^D$ of signature $(D,D)$, with bilinear form $p_L\cdot p_L'-p_R\cdot p_R'$~\cite{Narain1986,NarainSarmadiWitten1987}. With the same current normalization as~\eqref{eq:heisenberg-convention} its states have
\begin{equation}
h=\frac{p_L^2}{2}+N_L,\qquad
\bar h=\frac{p_R^2}{2}+N_R,\qquad
(p_L,p_R)\in\Gamma,\quad N_L,N_R\in\mathbb Z_{\ge0}.
\label{eq:narain-weight-convention}
\end{equation}
The evenness of $\Gamma$ ensures integer spin in the theory before orbifolding. We will use two special Narain lattices constructed from a positive-definite even unimodular lattice $L$~\cite{DMS1997,Wendland2000,AKN2025}
\begin{align}
\Gamma_{\mathrm N}(L)
&=\left\{\left(\frac{m+w}{\sqrt2},\frac{m-w}{\sqrt2}\right)\,\middle|\,m,w\in L\right\},\notag\\
\Gamma_{\mathrm{split}}(L)&=L\oplus L(-1),
\label{eq:narain-parent-points}
\end{align}
where $L(-1)$ has opposite bilinear form to that of $L$. We denote the first Narain theory by $\mathcal N(L)$ in which left and right charges are constrained by the common pair $(m,w)$, whereas the second has state space $V_L\otimes\overline{V_L}$ and independent left and right charges. Both theories have $c_L=c_R=D$.

In this section, $G$ denotes the finite internal symmetry group of the parent theory that will be gauged. In our lattice constructions, the symmetry action must be specified on both currents and charged fields, including the phases required to preserve their operator products. For a lattice VOA $V_L$, this amounts to lifting lattice isometries to VOA automorphisms. Explicitly, for a lattice isometry $\bar g \in O(L)$, a lift is an automorphism $g\in \operatorname{Aut}(V_L)$ acting on charged states as $g(e^\lambda)=\eta_g(\lambda)e^{{\bar g}\lambda}$. Writing the twisted group-algebra product as
\begin{equation}
	e^\lambda e^\mu=\epsilon(\lambda,\mu)e^{\lambda+\mu},
\end{equation}
preservation of the twisted group-algebra product requires~\cite{BK2004}
\begin{equation}
	\epsilon(\lambda,\mu)\eta_g(\lambda+\mu) = \epsilon({\bar g}\lambda,{\bar g}\mu) \eta_g(\lambda)\eta_g(\mu).
\end{equation}
For a Narain theory, the analogous compatibility condition applies to the cocycle of its full momentum lattice $\Gamma$, with the lattice action preserving the left- and right-moving stress tensors. Here $G$ is the finite group generated by these specified quantum symmetries. We write $V^G$ for the fixed-point VOA, $\mathcal C/G$ for the full orbifold, $N_g$ for the order of $g$ as an actual symmetry, and $\mathcal H_g$ for the unprojected $g$-twisted sector.

Gauging requires the anomaly class of the specified symmetry action to vanish,
\begin{equation}
	[\alpha]=0\in H^3(G,U(1))\cong H^4(G,\mathbb Z),
\end{equation}
where the coefficient actions are trivial~\cite{Bhardwaj:2017xup,Brown1982}. For a strongly rational holomorphic VOA $V$, the restriction of the anomaly $[\alpha_V]$ to the cyclic group generated by $g\in G$ can be determined from twisted conformal weights. For every finite-order $g$, there is a unique irreducible $g$-twisted module $V(g)$ up to isomorphism~\cite[Theorem~1.2]{DLM2000}. Let $\rho_g$ be its lowest conformal weight. For $g$ of order $N_g$, cyclic orbifold theory assigns the type~\cite[paragraph following Theorem~5.11]{vEMS2020}
\begin{equation}
	t_g=N_g^2\rho_g\pmod{N_g}\in\mathbb Z_{N_g}.
	\label{eq:cyclic-type}
\end{equation}
Under the identification $H^3(\langle g\rangle,U(1))\cong\mathbb Z_{N_g}$ with generator $g$ in the conventions of~\cite{GK2021}, the type $t_g$ represents the restricted anomaly $\operatorname{res}^{G}_{\langle g\rangle}[\alpha_V]$. In particular, this restriction vanishes if and only if $t_g=0$, which is equivalent to the chiral level-matching condition $N_g\rho_g\in\mathbb Z$~\cite{vEMS2020,KRR2026}. For a full CFT with $c_L=c_R$, the cyclic condition is $N_g(h-\bar h)\in\mathbb Z$ for every state in the unprojected $g$-twisted sector~\cite{LinShao2021}. We note that for both full CFTs and holomorphic VOAs, vanishing of the anomaly on every cyclic subgroup need not imply vanishing of the full group anomaly. When the full anomaly class vanishes, a choice of trivialization $\delta\beta=\alpha$ of a representative cocycle $\alpha$ linearizes the projective action of $C_G(g)$ on each $g$-twisted sector $\mathcal H_g$. Denoting the resulting linearized actions by $U_g: C_G(g) \rightarrow \operatorname{Aut}(\mathcal H_g)$, the orbifold state space is
\begin{equation}
\mathcal H_{\mathcal C/G}
=\bigoplus_{[g]\in\operatorname{Conj}(G)}P_g\mathcal H_g,\qquad
P_g=\frac1{|C_G(g)|}\sum_{k\in C_G(g)}U_g(k),
\label{eq:orbifold-state-space}
\end{equation}
where the sum runs over conjugacy classes~\cite{DHVW1985,DHVW1986,DVVV1989}. For our unitary full CFTs, the actions $U_g(k)$ are unitary, so each $P_g$ is an orthogonal projector and its image inherits the positive-definite inner product of $\mathcal H_g$.

We use finite gauging as a construction of correlators as well as of the state space. The specified unitary automorphisms determine invertible symmetry defects $\mathfrak D_g$ of the parent theory. Once the anomaly vanishes, junction phases can be chosen so that the multiplication on $\bigoplus_{g\in G}\mathfrak D_g$ is associative. With unit-norm junctions, its adjoint divided by $|G|$ splits multiplication, and orientation reversal identifies $\mathfrak D_g$ with the dual of $\mathfrak D_{g^{-1}}$. These data form the symmetric separable defect algebra used in finite orbifolding. Inserting its networks defines local correlators compatible with sewing, while its adjoint structure and the orthogonal projectors above preserve unitarity~\cite[Secs.~4.3 and 5.3.1]{Bhardwaj:2017xup} and~\cite[Secs.~2.1--2.2]{DLWW2024}. Thus our use of anomaly-free gauging includes the operator products and their consistency, beyond a modular-invariant partition function.

To exclude low-dimension states from the twisted sectors in our constructions, we use a lower bound determined solely by the symmetry action on the free-boson currents. For an orthogonal matrix $R \in O(D)$ acting on one chiral set of $D$ free-boson currents with eigenvalues $e^{2\pi i\theta_a}$, $0\le\theta_a<1$, the oscillator contribution to the twisted ground-state conformal weight is~\cite{DFMS1987,KRR2026}
\begin{equation}
\rho_{\mathrm{osc}}(R)
=
\frac14\sum_{a=1}^D\theta_a(1-\theta_a)
\label{eq:lattice-twist-energy}
\end{equation}
where the sum counts all eigenvalues with multiplicity. Let $R_L(g)$ and $R_R(g)$ denote the actions on the left- and right-moving currents. Non-negativity of zero-mode energies and additional oscillator excitations gives $h+\bar h\ge\rho_{\mathrm{osc}}(R_L(g))+\rho_{\mathrm{osc}}(R_R(g))$ throughout the $g$-twisted sector~\cite{BK2004,KRR2026}. Projections can remove states but cannot lower this bound.

In the subsequent constructions we will distinguish genuinely full constructions from tensor-product ones, for which we call a full CFT nonfactorizing if it admits no decomposition into nontrivial CFTs with additive stress tensors, in particular no decomposition into purely holomorphic and antiholomorphic factors.

\section{\label{sec:CFTs}Quadratic-Residue Orbifolds with Virasoro Primary Gap Three}

In this section we construct a family of 2d full CFTs with Virasoro primary gap $\Delta_{\mathrm{gap}}=3$. Such CFTs are obtained from orbifolding parent CFTs associated with even unimodular lattices constructed using the quadratic residue (QR) codes by $\operatorname{SL}_2(p)\times\mathbb Z_2$. In subsection~\ref{subsec:QR_Quotient}, we show that lifted $\operatorname{SL}_2(p)\times\mathbb Z_2$-action on the parent CFT is anomaly-free. In subsection~\ref{subsec:QR_Spectrum} we compute the spectrum of the orbifold theory and show that all theories $\mathcal C_{\mathrm {QR}}/(\operatorname{SL}_2(p)\times\mathbb Z_2)$ in this family have Virasoro primary gap three.

\subsection{\label{subsec:QR_Quotient}Anomaly-free \texorpdfstring{$\operatorname{SL}_2(p)\times \mathbb Z_2$}{SL2(p) x Z2} action on \texorpdfstring{$\mathcal C_{\mathrm {QR}}$}{the QR parent}}

The Paley matrix $W$ enters the construction twice. Over $\mathbb F_3$, its$(-1)$-eigenspace is the code from which we build the parent lattice. Over $\mathbb R$, $W$ is skew-symmetric and is preserved by the signed $\operatorname{SL}_2(p)$ action. In subsection~\ref{subsec:QR_Spectrum} we show that the corresponding oscillator state $\sum_{i,j}W_{ij}a^i_{-2}a^j_{-1}\ket{0}$ spans the space of orbifold primaries of weights $(3,0)$.

\paragraph{Choice of parent lattice}
Let $p\ge71$ be a prime with $p\equiv23\pmod{24}$ and set $D=p+1$. We label the standard basis $\{e_x\}$ of $\mathbb R^D$ by the points $x$ of the projective line $\mathbb P^1(\mathbb F_p)=\mathbb F_p\cup\{\infty\}$, and we choose the representatives $v_x=(x,1)^T$ and $v_\infty=(1,0)^T$ of the corresponding lines in $\mathbb F_p^2$. Let $\chi_p$ be the quadratic character of $\mathbb F_p^\times$, extended by $\chi_p(0)=0$. The Paley matrix~\cite{Paley1933,deLauneyStafford2005,Chapman2001Conference}
\begin{equation}
	W_{xy}=\chi_p(\det(v_x,v_y)),\qquad W^T=-W,\qquad W^2=-pI_D,
	\label{eq:QR-Paley}
\end{equation}
is the skew-symmetric conference matrix of order $D$. Indeed, $W^T=-W$ follows from $\chi_p(-1)=-1$, every row of $W$ has exactly $p$ nonzero entries, and the off-diagonal entries of $WW^T$ vanish by the character sum $\sum_{u\in\mathbb F_p}\chi_p\bigl((u-x)(u-y)\bigr)=-1$ for $x\ne y$. Explicitly, $W_{xy}=\chi_p(x-y)$ for $x,y\in\mathbb F_p$, while $W_{x\infty}=-1$ and $W_{\infty y}=1$.

Since $p\equiv2\pmod3$, \eqref{eq:QR-Paley} gives $W^2\equiv I_D\pmod3$, so $\mathbb F_3^D=\ker(W-I_D)\oplus\ker(W+I_D)$. We set
\begin{equation}
	C_p=\ker(W+I_D)\subset\mathbb F_3^D.
	\label{eq:QR-code}
\end{equation}
If $Wu=\epsilon u$ and $Wv=\epsilon v$ with $\epsilon=\pm1$, then \eqref{eq:QR-Paley} gives $\epsilon\,u\cdot v=(Wu)\cdot v=-u\cdot(Wv)=-\epsilon\,u\cdot v$, hence $u\cdot v=0$. Thus both eigenspaces are totally isotropic, so both have dimension $D/2$, and $C_p$ is a self-dual code. Every row sum of $W$ is congruent to $-1$ modulo $3$, so the all-ones word $(1,\ldots,1)$ lies in $C_p$. Moreover, $C_p$ is monomially equivalent to the extended ternary quadratic-residue code of length $D$. Indeed, by the explicit form of $W$, the translations $x\mapsto x+b$ and the dilations $x\mapsto a^2x$ of $\mathbb P^1(\mathbb F_p)$, which fix $\infty$, commute with $W$, so they preserve $C_p$. Since $C_p$ is self-dual, no nonzero codeword is supported on $\{\infty\}$ alone. Hence deleting the coordinate $\infty$ maps $C_p$ isomorphically onto a cyclic code of length $p$ and dimension $(p+1)/2$ that is invariant under the dilations. Its set of zeros is a union of orbits of the nonzero squares acting on $\mathbb F_p$, and it has $(p-1)/2$ elements, so it consists of the quadratic residues or of the nonresidues. Thus the deleted code is an odd-like QR code, and $C_p$ is its extension~\cite[Secs.~14.2--14.4]{Mueller2024Coding}.

Applying Construction A to $C_p$ gives the odd unimodular lattice
\begin{equation}
	M_p=\frac1{\sqrt3} \{x\in\mathbb Z^D:x\bmod3\in C_p\}
\end{equation}
with sublattice $M_{p,0}=\{v\in M_p:v^2\in2\mathbb Z\}$ and characteristic $\kappa=\sqrt3(1,\ldots,1)$ in $M_p$. Since $D\in24\mathbb Z\subset8\mathbb Z$, applying the standard neighbor construction results in the even unimodular lattice~\cite{BouyuklievaBouyuklievHarada2013}
\begin{equation}
	L=L^{\mathrm{QR}}_{p+1} = M_{p,0}\cup\left(\frac{\kappa}{2}+M_{p,0}\right).
	\label{eq:qr-lattice}
\end{equation}
Indeed, for $v=x/\sqrt3\in M_p$ one has $(\kappa,v)=\sum_i x_i\equiv v^2\pmod2$. Since $v^2\in 2\mathbb Z$, $\kappa/2$ pairs integrally with $M_{p,0}$ and has even norm $3D/4$. Since $[M_p:M_{p,0}]=2$ and $M_p$ is unimodular, $\det M_{p,0}=4$. Moreover, since $(\kappa/2,M_{p,0})\subset\mathbb Z$ and $(\kappa/2)^2=3D/4\in2\mathbb Z$, adjoining the coset $\kappa/2+M_{p,0}$ to $M_{p,0}$ gives an even integral index-two overlattice of $M_{p,0}$, hence
\begin{equation}
	\det L=\frac{\det M_{p,0}}{2^2}=1.
\end{equation}
Related quadratic-residue constructions using ideal lattices are described in~\cite{Chapman2001Conference}.

The minimum squared norm in $L$ is six. To see this, let $d_p$ denote the minimum distance of $C_p$. For $p=71$ one has $d_p=18$~\cite{CoppersmithSeroussi1984} and for every larger allowed prime, $p\ge167$. The square-root bound gives $(d_p-1)(d_p-2)+1\ge p$~\cite{Mueller2024Coding}, and the self-duality of $C_p$ over $\mathbb F_3$ implies that every codeword of $C_p$ has Hamming weight divisible by three hence $3\mid d_p$. Thus, since $d_p\le12$ would give $p\le111$, we must have $d_p\ge15$ for $p\ge167$. Now there are two cases. When $v=x/\sqrt3\in M_{p,0}$ has nonzero residue in $C_p$, we have $v^2\ge d_p/3\ge5$ hence $v^2\ge6$ by the evenness of $M_{p,0}$. When $v=x/\sqrt3\in M_{p,0}$ has zero residue in $C_p$, we have $v=\sqrt3\,y$ with $0\ne y\in\mathbb Z^D$ and $3y^2=v^2\in2\mathbb Z$, so again $v^2\ge6$. Thus every nonzero vector of $M_{p,0}$ has norm at least six, while every vector $y\in\kappa/2+M_{p,0}$ has norm at least $D/12\ge6$ since $y_i^2\ge \frac1{12}$ for each coordinate $y_i$ of $y$. Thus we have shown $\min(L)\ge6$. Note that $\sqrt3(e_i+e_j)\in M_{p,0}$ for $i\ne j$ has norm six, we conclude $\min(L)=6$.

Using~\eqref{eq:narain-parent-points}, we define the parent Narain CFT $\mathcal C_{\mathrm{QR}}=\mathcal N(L)$ by
\begin{equation}
    (p_L,p_R)=\left(\frac{m+w}{\sqrt2},\frac{m-w}{\sqrt2}\right),
    \qquad (m,w)\in L\oplus L,\qquad (c_L,c_R)=(D,D).
\end{equation}
The momentum and winding contribution to the scaling dimension is
\begin{equation}
    \Delta_{\mathrm{charge}}=\frac{p_L^2+p_R^2}{2}
    =\frac{|m+w|^2+|m-w|^2}{4}
    =\frac{|m|^2+|w|^2}{2}\ge3\qquad\text{if }(m,w)\ne(0,0).
\end{equation}
More generally, a state in the parent theory has
\begin{equation}
    \Delta=\Delta_{\mathrm{charge}}+N_L+N_R\ge\Delta_{\mathrm{charge}}.
\end{equation}
Therefore all states with $0<\Delta<3$ are neutral under both $U(1)^D$ current symmetries and consist entirely of oscillator excitations. Examples include $J_L^i$ and $J_R^j$ with dimension one and $J_L^iJ_R^j$ with dimension two.

\paragraph{Specify the symmetry action}

We first describe the geometric subgroup $K=\operatorname{SL}_2(p)$ as a faithful subgroup of $O(L)$. We then lift its action to $\mathcal N(L)$ and adjoin momentum--winding exchange to obtain the full group $G$. To prove that this action is anomaly-free, we show that the anomaly is determined by its two factor restrictions and then prove that both restrictions vanish.

Let
\begin{equation}
    K=\mathrm{SL}_2(p)=\left\{\begin{pmatrix}a&b\\c&d\end{pmatrix}:a,b,c,d\in\mathbb F_p,\ ad-bc=1\right\},
    \qquad G=K\times\langle t\rangle,\qquad t^2=1.
    \label{eq:QR-group}
\end{equation}
The natural action of $K$ on $\mathbb F_p^2$ induces a projective action on $\mathbb P^1(\mathbb F_p)$, and with the representatives $v_x$ chosen above we write
\begin{equation}
	kv_x=\lambda(k,x)v_{k\cdot x}.
\end{equation}
The cocycle relation for $\lambda$ and multiplicativity of the quadratic character $\chi_p$ of $\lambda(k,x)\in\mathbb F_p^\times$ lead to the signed permutation representation
\begin{equation}
	r(k)e_x=\chi_p(\lambda(k,x))e_{k\cdot x}.
	\label{eq:QR-signed-action}
\end{equation}
Since $\det(kv_x,kv_y)=\det(v_x,v_y)$, the definition of $\lambda$ gives $r(k)^TWr(k)=W$, namely $W$ commutes with $r(K)$. Hence these signed permutations preserve $C_p=\ker(W+I_D)$, and therefore also $M_p$ and $M_{p,0}$. Moreover, the product of the coordinate signs, $\prod_{x\in\mathbb P^1(\mathbb F_p)} \chi_p(\lambda(k,x))$, is a homomorphism $K\to\{\pm1\}$ which is trivial because $H^1(K,\mathbb Z_2)=\operatorname{Hom}(K,\mathbb Z_2)=0$ since $K$ is perfect. Thus $r(k)$ changes an even number of signs, so
\begin{equation}
	r(k)\frac{\kappa}{2}-\frac{\kappa}{2}\in M_{p,0},
\end{equation}
and hence $r(k)L=L$. Finally, the projective kernel is $\{\pm I_2\}$ while $r(-I_2)=-I_D$, so
\begin{equation}
	r:K\hookrightarrow O(L)
\end{equation}
is faithful.

We first define the action of $G$ on the momentum--winding labels by
\begin{equation}
k:(m,w)\longmapsto(r(k)m,r(k)w),
\qquad
t:(m,w)\longmapsto(w,m).
\end{equation}
These transformations preserve the Narain momentum lattice and they act on the left- and right-moving momenta by $(r(k),r(k))$ and $(I,-I)$, respectively. Here the crucial point is that $t$ reverses the right-moving currents while leaving the left-moving currents unchanged.

To obtain symmetries of the parent CFT, we must lift this lattice action to the vertex operators~\cite{BK2004,HarveyMoore2018}. For this we have the following theorem.

\begin{theorem}[Lifting the quadratic-residue symmetry]
\label{thm:QR-lift}
For every prime $p\ge71$ with $p\equiv23\pmod{24}$, let $L=L_{p+1}^{\mathrm{QR}}$ be the positive-definite even unimodular lattice associated with the extended ternary quadratic-residue code. The above action of $G=\mathrm{SL}_2(p)\times\langle t\rangle$ on its Narain momentum lattice admits a faithful lift
\begin{equation}
    \iota:G\hookrightarrow\operatorname{Aut}_{\mathrm{unitary}}(\mathcal N(L)).
\end{equation}
The lift of $t$ is an involution $\mathcal T$ commuting with the lifted $\mathrm{SL}_2(p)$ action, and $\iota(k,t^\varepsilon)$ acts on the left- and right-moving free-boson currents by $r(k)$ and $(-1)^\varepsilon r(k)$, respectively.
\end{theorem}

We prove the theorem by constructing the lift explicitly. For the lattice vertex operator $V_{m,w}$ with momentum label $m$ and winding label $w$, we have~\cite{HarveyMoore2018}:
\begin{equation}
	V_{m,w}(z,\bar z) =\ : e^{ip_LX_L(z) + ip_RX_R(\bar z)} : \widehat c_{m,w}
\end{equation}
where $\widehat c_{m,w}$ is a cocycle operator that supplies a phase through
\begin{equation}
	\widehat c_{m,w} \ket{m',w'} = \epsilon\bigl((m,w),(m',w')\bigr)\ket{m',w'} = (-1)^{(w,m')}\ket{m',w'}.
\end{equation}
The diagonal action of $k\in K$ preserves $\epsilon$ because $r(k)$ preserves the inner product. For the momentum--winding exchange $\mathcal T$, the phase $\eta_t(m,w)=(-1)^{(m,w)}$ satisfies
\begin{equation}
	\frac{\eta_t(x+y)}{\eta_t(x)\eta_t(y)} = (-1)^{(m,w')+(m',w)} = \frac{\epsilon(tx,ty)}{\epsilon(x,y)}, \qquad x = (m,w),y = (m',w')\in L\oplus L,
	\label{eq:QR-lift-cocycle}
\end{equation}
which ensures compatibility with the vertex-operator products. Thus we define the lifted operators $D_k$ and $\mathcal T$ by
\begin{equation}
	D_k V_{m,w}D_k^{-1}=V_{r(k)m,r(k)w}, \qquad \mathcal T V_{m,w}\mathcal T^{-1} =(-1)^{(m,w)}V_{w,m},
	\label{eq:GonV}
\end{equation}
together with the induced rotations of the left- and right-moving current modes
\begin{equation}
	\begin{split}
		&D_k a_{-n}(v)D_k^{-1}=a_{-n}(r(k)v),\ D_k\bar a_{-n}(v)D_k^{-1}=\bar a_{-n}(r(k)v),\\
		&\mathcal T a_{-n}(v)\mathcal T^{-1}=a_{-n}(v),\ \mathcal T\bar a_{-n}(v)\mathcal T^{-1}=-\bar a_{-n}(v)
	\end{split}
	\label{eq:GonO}
\end{equation}
where $a_{-n}(v) = \sum_{i=1}^{D} v_i a_{-n}^i$ for $v \in \mathbb R^D$. These lifted operators obey
\begin{equation}
	D_kD_\ell=D_{k\ell}, \qquad \mathcal T^2=1, \qquad D_k\mathcal T=\mathcal T D_k
	\label{eq:QR-lift-relations}
\end{equation}
Indeed, $D_kD_\ell=D_{k\ell}$ since the lift of $K$ carries no phases, $\mathcal T^2V_{m,w}\mathcal T^{-2}=(-1)^{(m,w)+(w,m)}V_{m,w}=V_{m,w}$, and $D_k\mathcal T=\mathcal TD_k$ since $(r(k)m,r(k)w)=(m,w)$. In the language of~\cite[Sec.~3.1]{MollerScheithauer2019}, the phase $\eta_t$ is trivial on the fixed sublattice $\{(\lambda,\lambda):\lambda\in L\}$ of $t$ because $L$ is even, so $\mathcal T$ is a standard lift of $t$. Moreover, for $\gamma=(m,w)$ the Narain pairing gives $\langle\gamma,t\gamma\rangle=m^2+w^2\in2\mathbb Z$, which is precisely the condition under which a standard lift of an involution does not suffer order doubling~\cite[Sec.~3.1]{MollerScheithauer2019}, see also~\cite[Eq.~(2.17)]{HarveyMoore2018}. It is also easy to check that the lifted operators preserve the vacuum, the stress tensor $T(z) = \frac{1}{2}\sum_{i=1}^{D} : J_L^i(z)J_L^i(z) :$ and its right-handed counterpart, and the positive Hilbert-space inner product. Consequently,
\begin{equation}
	\iota:G\longrightarrow\operatorname{Aut}_{\mathrm{unitary}}(\mathcal N(L)), \qquad \iota(k,t^\varepsilon)=D_k\mathcal T^\varepsilon
\end{equation}
is a homomorphism, where $\operatorname{Aut}_{\mathrm{unitary}}(\mathcal N(L))$ is the group of unitary automorphisms of the full Narain CFT with target torus $L\otimes_{\mathbb Z}\mathbb R /(2\pi L)$. The map $\iota$ is faithful because $(k,t^\varepsilon)\in G$ acts on the left- and right-moving oscillator spaces as $(r(k),(-1)^\varepsilon r(k))$, which becomes identity only when $k=1$ and $\varepsilon=0$. Thus the abstract group $G$ acts faithfully by unitary automorphisms of the parent CFT.

We note that the construction above uses only the evenness and unimodularity of $L$ and the fact that $r(k)$ acts by lattice isometries. Hence, for every positive-definite even unimodular lattice $L$ and every faithful representation $r:K\to O(L)$ of a finite group $K$, the formulas~\eqref{eq:GonV} and~\eqref{eq:GonO} define a faithful unitary action of $K\times\langle t\rangle$ on $\mathcal N(L)$.

\paragraph{Prove that it is anomaly-free}
Now we need to establish that the anomaly of the $G$-action vanishes, so that it can be gauged. For this we have the following theorem.

\begin{theorem}[Anomaly cancellation for quadratic-residue orbifolds]
\label{thm:QR-anomaly}
For every prime $p\ge71$ with $p\equiv23\pmod{24}$, the action of $G=\mathrm{SL}_2(p)\times\langle t\rangle$ on $\mathcal N(L_{p+1}^{\mathrm{QR}})$ defined by~\eqref{eq:GonV} and~\eqref{eq:GonO} has trivial anomaly,
\begin{equation}
    [\alpha]=0\in H^3(G,U(1)).
\end{equation}
It therefore admits consistent finite gauging, with $(c_L,c_R)=(p+1,p+1)$.
\end{theorem}

The same question arises for the Narain parents of sections~\ref{sec:P48q} and~\ref{sec:T512}, and the argument does not depend on the particular lattice. We therefore prove Theorem~\ref{thm:QR-anomaly} as a special case of the following general result.

\begin{theorem}[Anomaly cancellation for Narain examples]
\label{thm:Narain-anomaly}
Let $L$ be a positive-definite even unimodular lattice of rank $D$, let $K$ be a finite perfect group with a faithful representation $r:K\to O(L)$, and let $G=K\times\langle t\rangle$ act on $\mathcal N(L)$ by~\eqref{eq:GonV} and~\eqref{eq:GonO}. Then the anomaly of this action is trivial,
\begin{equation}
    [\alpha]=0\in H^3(G,U(1)),
\end{equation}
and therefore the $G$-action admits consistent finite gauging.
\end{theorem}

We give the proof of Theorem~\ref{thm:Narain-anomaly} in Appendix~\ref{app:Narain-anomaly}, and only summarize its structure here. Since $K$ is perfect, the K\"unneth theorem splits $H_3(G,\mathbb Z)$ into $H_3(K,\mathbb Z)$, $H_3(\mathbb Z_2,\mathbb Z)$ and a mixed summand $H_2(K,\mathbb Z)\otimes\mathbb Z_2$. The restriction of $[\alpha]$ to $K$ vanishes because $\mathcal N(L)$ admits a simple Dirichlet boundary condition that is invariant under the lifted $K$-action. The restriction to $\langle t\rangle$ vanishes by level matching in the unprojected $\mathcal T$-twisted sector $\mathcal H_{\mathcal T}$, whose spectrum is
\begin{equation}
    h=\frac{\lambda^2}{4}+N_L,\qquad
    \bar h=\frac{D}{16}+N_R^{\mathrm{tw}},
    \qquad \lambda\in L,\quad
    N_L\in\mathbb Z_{\ge0},\quad
    N_R^{\mathrm{tw}}\in\tfrac12\mathbb Z_{\ge0},
    \label{eq:T-spectrum_Narain}
\end{equation}
so that $2(h-\bar h)\in\mathbb Z$ follows from the evenness of $L$ and $D\in8\mathbb Z$. Finally, the mixed summand is detected by the projective action of $K$ on $\mathcal H_{\mathcal T}$, which is trivial because $\mathcal H_{\mathcal T}$ has a unique ground state, of weights $(0,D/16)$.

Theorem~\ref{thm:QR-anomaly} now follows directly. Indeed, $K=\mathrm{SL}_2(p)$ is superperfect for every prime $p\ge5$~\cite[p.~306]{Hillman2002}, in particular perfect, and $r:K\hookrightarrow O(L)$ is faithful. For superperfect $K$ the mixed summand is absent since $H_2(K,\mathbb Z) = 0$, so only the two restrictions have to be checked. Applying Theorem~\ref{thm:Narain-anomaly} with $L=L^{\mathrm{QR}}_{p+1}$ gives $[\alpha]=0$. Therefore we conclude that $G$ can be consistently gauged.

\subsection{\label{subsec:QR_Spectrum}Spectrum of the \texorpdfstring{$\mathcal C_{\mathrm {QR}}/(\operatorname{SL}_2(p)\times \mathbb Z_2)$}{QR} orbifold}

After consistently gauging $G$, the Hilbert space of the orbifold theory is~\cite{DHVW1985,DVVV1989}
\begin{equation}
	\mathcal H_{\mathrm{orb}} = \bigoplus_{[g]\subset G} \mathcal H_g^{\,C_G(g)} = \underbrace{\mathcal H_e^G}_{\text{invariant untwisted states}} \oplus \underbrace{\bigoplus_{[g]\ne[e]} \mathcal H_g^{C_G(g)}}_{\text{projected nonidentity twisted sectors}}.
\end{equation}
where $\mathcal H_g$ is the unprojected $g$-twisted sector and $C_G(g)$ is the centralizer of $g$. Recall that in $\mathcal N(L)$ since the lattice has minimum squared norm $6$, we have
\begin{equation}
	\Delta=\frac{m^2+w^2}{2}+N_L+N_R\ge3 \qquad\text{if }(m,w)\ne(0,0)
	\label{eq:DeltaBound_QR}
\end{equation}
Thus asking $\Delta < 3$ forces $m = w = 0$. Since for $z = -I_2 \in K$ we have
\begin{equation}
	\begin{split}
		D_z &: (a_{-n}^i,\bar a_{-n}^i) \longmapsto(-a_{-n}^i,-\bar a_{-n}^i),\\
       \mathcal T &: (a_{-n}^i,\bar a_{-n}^i) \longmapsto(a_{-n}^i,-\bar a_{-n}^i),\\
       D_z\mathcal T &: (a_{-n}^i,\bar a_{-n}^i) \longmapsto(-a_{-n}^i,\bar a_{-n}^i),
\end{split}
\end{equation}
an invariant state with $m = w = 0$ must contain an even number of creation operators in each chiral sector. This leaves $a_{-1}^i a_{-1}^j|0\rangle$ and $\bar a_{-1}^i\bar a_{-1}^j|0\rangle$ for states possibly with $\Delta = 2$. However, for a state $\sum_{ij} A_{ij}a_{-1}^ia_{-1}^j\ket{0}$ to be $K$-invariant, the symmetric tensor $A$ must be proportional to $\delta_{ij}$. Moreover, for $p\equiv3\pmod4$, the two irreducible constituents of $r_{\mathbb C}$ are complex conjugate, so the underlying real representation $r$ is irreducible. The real form of Schur's lemma therefore gives
\begin{equation}
	\bigl(\operatorname{Sym}^2r_{\mathbb C}^*\bigr)^K=\mathbb C\delta.
\end{equation}
Hence the only invariant left-moving state at weight two is $\sum_i a_{-1}^ia_{-1}^i\ket{0}=2L_{-2}\ket{0}$, which is a vacuum descendant since $L_2L_{-2}|0\rangle = \frac{c_L}{2}|0\rangle = \frac D2|0\rangle\ne 0$. Therefore we have $(\mathcal H_e^G)_{\Delta < 3} = \operatorname{span}_{\mathbb C}\{ \ket{0}, L_{-2}\ket{0}, \overline L_{-2}\ket{0} \}$.

For the twisted sectors $\mathcal H_g^{C_G(g)}$, we look at $\mathcal H_g$ first before the $C_G(g)$ projection. Since zero-mode contributions to the conformal weights are nonnegative, oscillator ground-state energies give lower bounds in these sectors. Recall that by~\eqref{eq:lattice-twist-energy}, for an orthogonal transformation $R$ with eigenvalues $e^{2\pi i\theta_a}$ where $0\le\theta_a<1$, its chiral oscillator ground-state weight is
\begin{equation}
	\rho_{\mathrm{osc}}(R) =\frac14\sum_{a=1}^{D}\theta_a(1-\theta_a).
\end{equation}
Thus for all states in $\mathcal H_g$ we have
\begin{equation}
	\Delta \geq \rho_{\mathrm{osc}}(R_L(g)) + \rho_{\mathrm{osc}}(R_R(g)). 
\end{equation}
On one hand, for twists containing $\mathcal T$, say $g = D_k\mathcal T$, we have $R_L(g) = r(k)$ and $R_R(g) = -r(k)$. We write $f(\theta)=\theta(1-\theta)$. For $0\le\theta<1$, since multiplying an eigenvalue by $-1$ shifts the corresponding $\theta$ by $1/2$, we have:
\begin{equation}
	f(\theta) + f(\{\theta + 1/2\}) \geq \frac{1}{4}
\end{equation}
where $\{\cdot\}$ denotes the fractional part. Thus, summing over the $D$ oscillators we have:
\begin{equation}
	\Delta\ge \rho_{\mathrm{osc}}(r(k)) + \rho_{\mathrm{osc}}(-r(k)) \ge\frac D{16}\ge\frac92>3.
\end{equation}

On the other hand, for $g=D_k$ with $k\ne1$, since both sides have the same twist, we have
\begin{equation}
	\Delta\ge2\rho_{\mathrm{osc}}(r(k)).
\end{equation}
Thus the problem becomes finding the eigenvalues of $r(k)$ for all $k\in K$. The representation~\eqref{eq:QR-signed-action} acts by signed permutations in the orthonormal coordinate basis of $L\otimes_{\mathbb Z}\mathbb R$. For a cycle of length $\ell$, we write
\begin{equation}
	r(k)e_1=s_1e_2,\quad \ldots,\quad r(k)e_\ell=s_\ell e_1,
\qquad s_j=\pm1.
\end{equation}
where relevant sign is the product around the entire cycle $S=s_1s_2\cdots s_\ell$. Individual $s_i$'s depend on the chosen orientations of the basis vectors, but $S$ does not. Hence we have $r(k)^\ell e_i = S e_i$ and the eigenvalue of $r(k)$ must satisfy $\lambda^\ell = S$. Therefore, writing $\lambda = e^{2\pi i \theta}$ with $\theta\in [0,1)$, we have:
\begin{equation}
	\theta = \begin{cases}
		\frac{j}{\ell}, & S = 1\\
		\frac{j + 1/2}{\ell}, & S = -1
	\end{cases}
	\qquad j = 0,\cdots,\ell-1
\end{equation}
Plugging the eigenvalues into~\eqref{eq:lattice-twist-energy}, we have
\begin{equation}
	\begin{split}
		\rho_+(\ell) &= \frac14\sum_{j=0}^{\ell-1} \frac{j}{\ell}\left(1-\frac{j}{\ell}\right) = \frac{\ell^2-1}{24\ell}, \\
		\rho_-(\ell) &= \frac14\sum_{j=0}^{\ell-1} \frac{j+\frac12}{\ell} \left(1-\frac{j+\frac12}{\ell}\right) = \frac{2\ell^2+1}{48\ell}
	\end{split}
\end{equation}
for $S = 1$ and $S = -1$, respectively. Therefore, denote by $c_\ell^+(k)$ and $c_\ell^-(k)$ the numbers of positive and negative cycles of length $\ell$, we have
\begin{equation}
	\rho_{\mathrm{osc}}(r(k)) = \sum_{\ell\ge1} \left( c_\ell^+(k)\rho_+(\ell) + c_\ell^-(k)\rho_-(\ell) \right) = \sum_{\ell\ge1} \left( c_\ell^+(k)\frac{\ell^2-1}{24\ell} + c_\ell^-(k)\frac{2\ell^2+1}{48\ell} \right).
	\label{eq:RhoCycles_QR}
\end{equation}

We only need a lower bound on $\rho_{\mathrm{osc}}(r(k))$ for $k\ne1$. From the explicit values of $\rho_\pm$,
\begin{equation}
	\rho_+(1)=0,\qquad
	\rho_+(\ell)\ge\frac{\ell}{27}\quad(\ell\ge3),\qquad
	\rho_-(\ell)>\frac{\ell}{24}\quad(\ell\ge1),
	\label{eq:RhoPerCycle_QR}
\end{equation}
so every cycle of $r(k)$ contributes at least $1/27$ per basis vector, except positive cycles of length one, which contribute nothing, and positive cycles of length two, which contribute $\rho_+(2)=1/16$. We show that the latter never occur and that there are at most two of the former.

Suppose first that $x\ne y$ form a cycle of length two, so that $kv_x=\lambda v_y$ and $kv_y=\mu v_x$ with $\lambda=\lambda(k,x)$ and $\mu=\lambda(k,y)$. In the basis $(v_x,v_y)$ of $\mathbb F_p^2$ the matrix of $k$ is $\left(\begin{smallmatrix}0&\mu\\ \lambda&0\end{smallmatrix}\right)$, so $\det k=1$ forces $\lambda\mu=-1$. The sign of the cycle is therefore $S=\chi_p(\lambda)\chi_p(\mu)=\chi_p(-1)=-1$, since $p\equiv3\pmod4$ hence $-1$ is not a square in $\mathbb F_p$. Hence $c_2^+(k)=0$ for every $k\in K$.

Next, $x$ is fixed by the projective action of $k$ precisely when $v_x$ is an eigenvector of $k$. If $k\ne\pm I_2$ then $k$ is not scalar and has at most two eigenlines in $\mathbb F_p^2$, so $c_1^+(k)\le2$. For $k=z$ we have $r(z)=-I_D$ and hence $c_1^+(z)=0$. Thus $c_1^+(k)\le2$ for every $k\ne1$.

Since $\sum_\ell\ell\,\bigl(c_\ell^+(k)+c_\ell^-(k)\bigr)=D$, combining these facts with~\eqref{eq:RhoCycles_QR} and~\eqref{eq:RhoPerCycle_QR} gives
\begin{equation}
	\rho_{\mathrm{osc}}(r(k))\ge\frac{D-c_1^+(k)}{27}\ge\frac{D-2}{27}\qquad\text{for every }k\ne1,
	\label{eq:RhoMin_QR}
\end{equation}
so every state of a geometric twisted sector $\mathcal H_{D_k}$ with $k\ne1$ obeys $\Delta\ge2(D-2)/27$. Combining this with the bound $\Delta\ge D/16$ for the sectors twisted by $D_k\mathcal T$, and using $D/16\le2(D-2)/27$ for $D\ge13$, we conclude that
\begin{equation}
	\Delta\ge\frac{D}{16}\ge\frac92>3 \qquad\text{for every state of }\mathcal H_g\text{ with }[g]\ne[e],
	\label{eq:TwistedBound_QR}
\end{equation}
where for the numerical bound we use $D=p+1\ge72$. Since the projections only remove states, the same bound holds for $\mathcal H_g^{C_G(g)}$. Together with
\begin{equation}
	(\mathcal H_e^G)_{\Delta<3}=\operatorname{span}_{\mathbb C}\{\ket{0},L_{-2}\ket{0},\overline L_{-2}\ket{0}\}
\end{equation}
this shows that the only states of $\mathcal H_{\mathrm{orb}}$ with $\Delta<3$ are the vacuum and the two stress tensors. Hence
\begin{equation}
	\Delta_{\mathrm{gap}}\!\left(\mathcal C_{\mathrm{QR}}/G\right)\ge3.
\end{equation}

It remains to exhibit primaries at $\Delta=3$. By~\eqref{eq:TwistedBound_QR} all of them lie in $\mathcal H_e^G$, so we can enumerate them completely. A charged state has $\Delta=3$ only if $\frac{m^2+w^2}{2}=3$ and $N_L=N_R=0$, which by $\min(L)=6$ forces $(m,w)=(\mu,0)$ or $(0,\mu)$ with $\mu^2=6$. Such states have $(h,\bar h)=(3/2,3/2)$ and carry no oscillators, so they are Virasoro primaries. Fixing $\mu$ with $\mu^2=6$, we set
\begin{equation}
	\mathcal O_\mu = \sum_{k\in K}\left( \ket{r(k)\mu,0} + \ket{0,r(k)\mu} \right) = \left|\operatorname{Stab}_K(\mu)\right| \sum_{\nu\in K\mu} \left( \ket{\nu,0} + \ket{0,\nu} \right).
	\label{eq:OrbitSum_QR}
\end{equation}
By~\eqref{eq:GonV} the operators $D_k$ permute the summands without phases, and $\mathcal T\ket{\mu,0}=\ket{0,\mu}$ because $(\mu,0)=0$, so $\mathcal O_\mu$ is $G$-invariant. The second form in~\eqref{eq:OrbitSum_QR} is a sum of distinct basis vectors with equal positive coefficients, so $\mathcal O_\mu\ne0$. Hence $\mathcal O_\mu\in\mathcal H_e^G$ is a scalar Virasoro primary with $\Delta=3$, and these orbit sums span the $(3/2,3/2)$ primaries of $\mathcal H_{\mathrm{orb}}$, one for each $K$-orbit of norm-six vectors of $L^{\mathrm{QR}}_{p+1}$.

The neutral states with $\Delta=3$ are more restricted.  Invariance under $z$ and $t$ requires an even number of creation operators in each chiral sector, so the only possibilities are
\begin{equation}
	P(A)=\sum_{i,j}A_{ij}a^i_{-2}a^j_{-1}|0\rangle
\end{equation}
at $(h,\bar h)=(3,0)$ and their right-moving counterparts. Using $[L_n,a^i_m]=-m\,a^i_{n+m}$ it is not hard to show that $P(A)$ is primary precisely when $A^T=-A$, while $K$-invariance of $P(A)$ is equivalent to $A\in\operatorname{End}_K(r_{\mathbb C})$.

To write down $A$ concretely, we note that $r_{\mathbb C}$ decomposes into two inequivalent irreducible constituents~\cite{Mueller2024Coding}, so by Schur's lemma we have $\dim\operatorname{End}_K(r_{\mathbb C})=2$. Since the invariant symmetric tensors were already shown above to be spanned by $I$, the invariant antisymmetric tensors must form another one-dimensional space. The Paley matrix $W$ of~\eqref{eq:QR-Paley} is antisymmetric, and we showed in subsection~\ref{subsec:QR_Quotient} that it commutes with $r(K)$, therefore it has to span this space. Thus
\begin{equation}
	P_3:=P(W).
	\label{eq:QR-P3}
\end{equation}
spans the $(3,0)$ primaries.  Moreover
\begin{equation}
	\|P_3\|^2=2pD>0.
\end{equation}
The analogous right-moving state spans the $(0,3)$ primaries.

This completes the determination of the spectrum below and at the gap. Dirichlet's theorem gives infinitely many primes $p\equiv23\pmod{24}$, since $23$ and $24$ are coprime. Their distinct central charges make the resulting CFTs inequivalent. We summarize the construction as follows.

\begin{theorem}[Gap of the quadratic-residue orbifolds]
\label{thm:QR-gap}
Let $p\ge71$ be a prime with $p\equiv23\pmod{24}$, let $\mathcal C_{\mathrm{QR}}=\mathcal N(L^{\mathrm{QR}}_{p+1})$, and let $G=\mathrm{SL}_2(p)\times\langle t\rangle$ act as in~\eqref{eq:GonV} and~\eqref{eq:GonO}. Then the orbifold $\mathcal C_{\mathrm{QR}}/G$ is a compact unitary bosonic full CFT with a unique vacuum and $(c_L,c_R)=(p+1,p+1)$, and it has
\begin{equation}
	\Delta_{\mathrm{gap}}\!\left(\mathcal C_{\mathrm{QR}}/G\right)=3 .
\end{equation}
Its space of primaries of dimension three is spanned by the orbit sums $\mathcal O_\mu$ of weights $(3/2,3/2)$, one for each $\mathrm{SL}_2(p)$-orbit of norm-six vectors of $L^{\mathrm{QR}}_{p+1}$, together with $P_3$ and $\overline P_3$ of weights $(3,0)$ and $(0,3)$. Moreover $\mathcal C_{\mathrm{QR}}/G$ is nonfactorizing.
\end{theorem}

Unitarity and compactness follow from Theorems~\ref{thm:QR-lift} and~\ref{thm:QR-anomaly}: the twisted sectors of a Narain theory carry positive-definite inner products, the projectors $P_g$ of~\eqref{eq:orbifold-state-space} are orthogonal because the linearized actions $U_g(k)$ are unitary, and the charge and oscillator gradings keep the multiplicity below any fixed dimension finite. Uniqueness of the vacuum follows from~\eqref{eq:TwistedBound_QR} together with $\left(\mathcal H_e^G\right)_{\Delta=0}=\mathbb C\ket{0}$.

It remains to show that $\mathcal C_{\mathrm{QR}}/G$ is nonfactorizing. Suppose
\begin{equation}
	\mathcal C_{\mathrm{QR}}/G=\mathcal A\otimes\mathcal B
\end{equation}
with additive stress tensors. If both factors have positive left central charge, then
\begin{equation}
	X=c_{L,\mathcal B}T_{\mathcal A} - c_{L,\mathcal A}T_{\mathcal B}
\end{equation}
is a nonzero Virasoro primary of weight $(2,0)$ and the analogous statement holds on the right. Since the orbifold has primary gap three, on each chiral side at most one factor can have nonzero central charge. If the same factor, say $\mathcal B$, has vanishing left and right central charges $c_{L,\mathcal B} = c_{R,\mathcal B} = 0$, unitarity and uniqueness of the vacuum of $\mathcal C_{\mathrm {QR}}/G$ force $\mathcal B$ to be trivial. Thus the only possible nontrivial factorization is
\begin{equation}
	\mathcal H_{\mathrm{orb}}=V\otimes\overline{V'},
\end{equation}
with $V$ a purely left-moving state space and $\overline{V'}$ a purely right-moving state space.

For such a factorization the primary spaces factorize as
\begin{equation}
	P_{h,\bar h}(\mathcal H_{\mathrm{orb}}) = P_h(V)\otimes P_{\bar h}(\overline{V'}).
\end{equation}
The gap-three condition gives $P_2(V)=P_2(\overline{V'})=0$, and hence $P_{2,2}(\mathcal H_{\mathrm{orb}})=0$. On the other hand,
\begin{equation}
	Q_{2,2} = \sum_{i,j} \left( a^i_{-1}a^j_{-1} - \frac{\delta^{ij}}{D}\sum_\ell a^\ell_{-1}a^\ell_{-1} \right) \left( \bar a^i_{-1}\bar a^j_{-1} - \frac{\delta^{ij}}{D}\sum_\ell \bar a^\ell_{-1}\bar a^\ell_{-1} \right)|0\rangle
\end{equation}
is a $G$-invariant $(2,2)$ Virasoro primary, and
\begin{equation}
	\|Q_{2,2}\|^2=2(D-1)(D+2)>0.
\label{eq:mixed-primary-norm}
\end{equation}
This contradicts $P_{2,2}(\mathcal H_{\mathrm{orb}})=0$, hence we prove that $\mathcal C_{\mathrm{QR}}/G$ is nonfactorizing.

\section{\label{sec:P48q}The \texorpdfstring{$P_{48q}$}{P48q} Orbifold and the Gap-Three Family}

The argument of section~\ref{sec:CFTs} also gives a full CFT with Virasoro primary gap three at $(c_L,c_R)=(48,48)$. The representation of $r:K=\operatorname{SL}_2(p)\longrightarrow O(D)$ on the currents is the representation~\eqref{eq:GonO} of~\eqref{eq:QR-signed-action} on the currents specialized to $p=47$, but the underlying even unimodular lattice must be chosen differently. Indeed, $p=47$ is a prime with $p\equiv23\pmod{24}$, so all ingredients of section~\ref{sec:CFTs} exist at $D=48$: the extended ternary QR code $C_{47}$, the odd unimodular lattice $M_{47}$ obtained from it by Construction A, its even sublattice $M_{47,0}$, and the signed action~\eqref{eq:QR-signed-action} of $\operatorname{SL}_2(47)$. The step that fails is the bound $\min(L)\ge6$ for the neighbor~\eqref{eq:qr-lattice}. There the vectors of the coset $\frac{\kappa}{2}+M_{p,0}$ have norm at least $D/12$, which is at least six only for $D\ge72$, and at $p=47$ the all-ones word of $C_{47}$ produces vectors of norm four in this coset, so the analog of~\eqref{eq:DeltaBound_QR} would only give $\Delta\ge2$ for charged states. However, $M_{47,0}$ has exactly one other even unimodular overlattice of index two, obtained by adjoining the coset $\frac{\kappa}{2}+u+M_{47,0}$ with $u\in M_{47}$ of odd norm, and this neighbor of $M_{47}$ is the extremal lattice $P_{48q}$ of minimum norm six~\cite[proof of Thm.~3.6]{Nebe2014}, see also~\cite[Sec.~5.2]{Chapman2001Conference}. It is preserved by the same signed action of $\operatorname{SL}_2(47)$, so the currents transform exactly as in section~\ref{sec:CFTs}, and only the allowed momentum--winding charges change. Moreover, unlike the case in section~\ref{sec:CFTs} where no state in any twisted sectors saturate the bound $\Delta=3$, in this case the lowest states in two nontrivial twisted sectors do reach $\Delta=3$. In this section we explain these differences and provide the lattice and twisted-sector data needed to complete the construction.

\subsection{The lattice and the anomaly-free action}

Let $P_{48q}$ be the extremal even unimodular lattice of rank $48$ with lattice data~\cite{Nebe2014,P48q}
\begin{equation}
    \min(P_{48q})=6,\qquad
    K_{48}=\operatorname{Aut}(P_{48q})\cong\operatorname{SL}_2(47),\qquad
    |K_{48}|=47(47^2-1)=103776.
    \label{eq:P48q-data}
\end{equation}
Here extremality means that the minimum norm attains the even-unimodular bound $2+2\lfloor48/24\rfloor=6$. The monomial action identified in~\cite{Nebe2014} gives the same current representation as~\eqref{eq:QR-signed-action} at $p=47$. In suitable orthonormal coordinates it takes the form
\begin{equation}
    r_{48}(k)e_x=\chi_{47}(\lambda(k,x))e_{k\cdot x},
    \qquad x\in\mathbb P^1(\mathbb F_{47}),
    \label{eq:P48q-current-action}
\end{equation}
with $v_x$ and $\lambda(k,x)$ as in~\eqref{eq:QR-signed-action}. In particular $r_{48}(-I_2)=-I_{48}$.

The lattice construction nevertheless requires a different choice of even unimodular neighbor from~\eqref{eq:qr-lattice} at $p=47$. By~\eqref{eq:QR-code} at $p=47$, the all-ones word $(1,\ldots,1)$ belongs to $C_{47}$, so $\frac{\kappa}{2}+M_{47,0}$ contains
\begin{equation}
    \frac{1}{2\sqrt3}(1,\ldots,1)
    =\frac{\kappa}{2}-\frac1{\sqrt3}(1,\ldots,1),
    \qquad
    \left|\frac{1}{2\sqrt3}(1,\ldots,1)\right|^2=4.
\end{equation}
Hence in this case $M_{47,0}\cup \left(\frac{\kappa}{2}+M_{47,0}\right)$ has minimum norm four~\cite{Nebe2014}, so it does not satisfy the minimum-six condition of the QR family in section~\ref{sec:CFTs}. We therefore choose $P_{48q}$, the other even unimodular neighbor of the same odd code lattice $M_{47}$, which has minimum norm six. Both $M_{47,0}\cup \left(\frac{\kappa}{2}+M_{47,0}\right)$ and $P_{48q}$ are preserved by the same orthogonal $\operatorname{SL}_2(47)$ action~\cite{Nebe2014}, so their associated Narain theories have the same current representation~\cite{BK2004}, though their allowed momentum--winding charges differ.

We define the Narain parent $\mathcal C_{48}^{\mathrm{parent}}=\mathcal N(P_{48q})$ using~\eqref{eq:narain-parent-points}, and let
\begin{equation}
    G_{48}=K_{48}\times\langle t\rangle,
    \qquad |G_{48}|=207552
\end{equation}
be the symmetry group to be gauged. The cocycle $\epsilon((m,w),(m',w'))=(-1)^{(w,m')}$ and the operators~\eqref{eq:GonV} and~\eqref{eq:GonO}, with $r$ replaced by $r_{48}$, give a faithful unitary action. The calculation in the proof of Theorem~\ref{thm:QR-lift}, in particular~\eqref{eq:QR-lift-cocycle}--\eqref{eq:QR-lift-relations}, uses only the integrality of the lattice pairing and the fact that $r_{48}$ acts by lattice isometries, so it applies without modification.

The anomaly cancellation also carries over directly. Since $K_{48}\cong\operatorname{SL}_2(47)$ is superperfect~\cite[p.~306]{Hillman2002}, in particular perfect, and $r_{48}$ is faithful, Theorem~\ref{thm:Narain-anomaly} applies with $L=P_{48q}$. At $D=48$ the $\mathcal T$-twisted spectrum~\eqref{eq:T-spectrum_Narain} takes the form
\begin{equation}
    h=\frac{\lambda^2}{4}+N_L,\qquad
    \bar h=3+N_R^{\mathrm{tw}},\qquad
    \lambda\in P_{48q},\quad
    N_L\in\mathbb Z_{\ge0},\quad
    N_R^{\mathrm{tw}}\in\frac12\mathbb Z_{\ge0},
    \label{eq:P48q-T-spectrum}
\end{equation}
and since $P_{48q}$ is even, the level-matching condition
\begin{equation}
    2(h-\bar h)
    =\frac{\lambda^2}{2}+2N_L-6-2N_R^{\mathrm{tw}}
    \in\mathbb Z
\end{equation}
used in the proof of Theorem~\ref{thm:Narain-anomaly} holds explicitly. Thus $[\alpha]=0$ in $H^3(G_{48},U(1))$. Therefore $G_{48}$ can be consistently gauged.

\subsection{The spectrum below and at dimension three}

In the untwisted sector, states with nonzero charges have $\Delta\ge\min(P_{48q})/2=3$. The independent chiral sign reversals and the unique invariant symmetric form remove every neutral non-vacuum primary below three, exactly as in subsection~\ref{subsec:QR_Spectrum}, thus
\begin{equation}
    \left(\mathcal H_e^{G_{48}}\right)_{\Delta<3} = \operatorname{span}_{\mathbb C}\{\ket{0},L_{-2}\ket{0},\bar L_{-2}\ket{0}\}.
\end{equation}
At dimension three the invariant untwisted primaries consist of the charge-orbit sums~\eqref{eq:OrbitSum_QR} with $\mu\in P_{48q}$ of norm six and the two chiral oscillator primaries $P(W)$ and $\overline{P(W)}$. Here
\begin{equation}
    \begin{aligned}
    &W_{xy}=\chi_{47}(\det(v_x,v_y)),\qquad
    W^T=-W,\qquad W^2=-47I_{48},\\
    &\|P(W)\|^2=2\cdot47\cdot48=4512.
    \end{aligned}
\end{equation}
In particular the untwisted sector already contains nonzero primaries at the desired gap.

For geometric twists involving only $D_k$, the argument leading to~\eqref{eq:RhoMin_QR} uses only $p\equiv3\pmod4$ and therefore applies verbatim at $p=47$, giving
\begin{equation}
    \rho_{\mathrm{osc}}(r_{48}(k))\ge\frac{46}{27}, \qquad \Delta\ge\frac{92}{27} > 3 \quad\text{in }\mathcal H_{D_k},\quad k\ne1.
\end{equation}
For a twist $D_k\mathcal T$, the left- and right-moving oscillator actions are $r_{48}(k)$ and $-r_{48}(k)$. Each eigenangle $\theta$ of $r_{48}(k)$ is therefore paired with $\{\theta+\frac12\}$, and
\begin{equation}
    \theta(1-\theta)
    +\left\{\theta+\frac12\right\}
     \left(1-\left\{\theta+\frac12\right\}\right)
    \ge\frac14.
\end{equation}
Consequently,
\begin{equation}
    \Delta\ge\rho_{\mathrm{osc}}(r_{48}(k)) + \rho_{\mathrm{osc}}(-r_{48}(k)) \ge\frac{48}{16}=3
    \label{eq:P48q-twisted-bound}
\end{equation}
where equality requires every eigenangle to be $0$ or $\frac12$, hence every eigenvalue of $r_{48}(k)$ is $1$ or $-1$. The faithfulness of $r_{48}$ then implies $k^2=1$, and the only possibilities in $\operatorname{SL}_2(47)$ are $k=1,z$. Thus only the $\mathcal T$ and $D_z\mathcal T$ sectors can contribute twisted states at dimension three.

There is exactly one such ground state in each sector before projection. Indeed, the trace~\eqref{eq:ZT1_Narain} of Appendix~\ref{app:Narain-anomaly} becomes
\begin{equation}
    Z_{\mathcal T,1}
    =\frac{\Theta_{P_{48q}}(\tau/2)}{\eta(\tau)^{48}}
    \overline{\left(\frac{\eta(\tau)}{\eta(\tau/2)}\right)^{48}}
    =q^{-2}\bar q\left(1+\text{terms of higher total weight}\right)
\end{equation}
whose leading coefficient is one with weights $(0,3)$ since $h-\frac{48}{24}=-2$ and $\bar h-\frac{48}{24}=1$. The $D_z\mathcal T$ trace exchanges the two chiralities and has a unique ground state of weights $(3,0)$. These ground states are Virasoro primaries because every positive Virasoro mode would lower their weight within the same twisted sector, which is impossible.

We note that both the $(3,0)$ and the $(0,3)$ ground states survive projection by the corresponding centralizer. Indeed, for either $g=\mathcal T$ or $g=D_z\mathcal T$, the twist is central in $G_{48}$ hence its one-dimensional ground-state space must carry a one-dimensional representation of $G_{48}$. The perfect subgroup $K_{48}$ acts trivially on this space since every one-dimensional representation of a perfect group is trivial, while $g$ acts as $e^{2\pi i(h-\bar h)}=1$. Since $G_{48}$ is generated by $K_{48}$ and $g$ in either case, both ground states are $G_{48}$-invariant.

We have therefore proved the following result.
\begin{theorem}[Gap of the rank-48 orbifold]
\label{thm:P48q-gap}
The orbifold
\begin{equation}
    \mathcal C_{48}=\mathcal N(P_{48q})/
    \bigl(\operatorname{SL}_2(47)\times\langle t\rangle\bigr)
\end{equation}
with the lifted action above is a compact unitary bosonic full CFT with a unique vacuum, $(c_L,c_R)=(48,48)$ and $\Delta_{\mathrm{gap}}=3$. Its dimension-three primaries are the norm-six charge-orbit sums at $(3/2,3/2)$, one untwisted and one twisted primary at $(3,0)$, and their right-moving counterparts at $(0,3)$. The theory is nonfactorizing.
\end{theorem}

Unitarity and compactness follow from the same finite gauging construction as for the QR family. The positive bound~\eqref{eq:P48q-twisted-bound} ensures that no new vacuum appears. The nonfactorization proof uses the invariant $Q_{2,2}$ and its norm~\eqref{eq:mixed-primary-norm} with $D=48$, hence is unchanged. Thus $P_{48q}$ extends the gap-three construction in section~\ref{sec:CFTs} to a smaller central charge, but differs from the QR family in two concrete ways. Namely, it uses the minimum-six lattice $P_{48q}$ that is not in the QR family, and its twisted sectors supply a second primary in each chiral weight-three space whereas for the QR family with $p\ge71$, the spaces of primaries of weights $(3,0)$ and $(0,3)$ are each one-dimensional, and all dimension-three primaries lie in the untwisted sector.

\section{\label{sec:T512}The \texorpdfstring{$T_{512}$}{T512} Orbifold with Virasoro Primary Gap Four}

The same Narain construction gives a full CFT with Virasoro primary gap four at $(c_L,c_R)=(512,512)$ by taking the tensor lattice $T_{512}:=E_8^{\otimes3}$. Unlike the QR and $P_{48q}$ cases, its geometric current representation has no invariant antisymmetric bilinear, so the weight-three oscillator primaries are absent. The geometric group is perfect, but its second homology can support a mixed anomaly with the momentum--winding exchange. This mixed anomaly vanishes by Theorem~\ref{thm:Narain-anomaly}, and we then determine the gap.

\subsection{\label{subsec:T512_Quotient}The lattice and the anomaly-free action}

Let $E_8$ have its standard normalization with minimum norm two, and set
\begin{equation}
    L=E_8\otimes_{\mathbb Z}E_8\otimes_{\mathbb Z}E_8,
    \qquad D=512,\qquad \min(L)=8.
    \label{eq:T512-lattice}
\end{equation}
The inner product on pure tensors is the product of the three $E_8$ inner products. The Gram matrix is therefore the tensor cube of an $E_8$ Gram matrix, so $L$ is positive definite, even and unimodular. By Theorem~7.1.1 in~\cite{Kitaoka1999}, every minimal vector of $L\otimes M$ is a pure tensor whenever $\min\{\operatorname{rank}L,\operatorname{rank}M\}\leq 43$. Applying it twice with the $E_8$ factor gives the minimum in~\eqref{eq:T512-lattice}.

Let $H=W(E_8)'\cong2.O_8^+(2)$ be the derived subgroup of the Weyl group, where $O_8^+(2)$ denotes the simple group in ATLAS notation~\cite{Atlas1985,Lubeck2021}. Moreover, since $H$ is quasisimple, it is perfect~\cite{Lubeck2021}. Hence the product group $H^{\times3}$ is also perfect and acts on $L$ by $(h_1,h_2,h_3)\mapsto h_1\otimes h_2\otimes h_3$ whose kernel is
\begin{equation}
	N=\{(\epsilon_1I_8,\epsilon_2I_8,\epsilon_3I_8): \epsilon_i=\pm1,\ \epsilon_1\epsilon_2\epsilon_3=1\} \cong\mathbb Z_2^2.
\end{equation}
Hence the faithful lattice action is
\begin{equation}
    K=H^{\times3}/N,\qquad r([h_1,h_2,h_3])=h_1\otimes h_2\otimes h_3.
    \label{eq:T512-current-action}
\end{equation} Since subgroup $N$ is normal, $K$ is perfect because a quotient of a perfect group by a normal subgroup is again perfect~\cite{Berrick1984}. Therefore $H_1(K,\mathbb Z)=0$ and, for coefficients acted trivially by $K$, $H^1(K,U(1))\cong\operatorname{Hom}(K_{\mathrm{ab}},U(1))=0$ as well~\cite{Brown1982}.

We take the parent $\mathcal N(L)$ of~\eqref{eq:narain-parent-points} and consider gauging
\begin{equation}
    G_{512}=K\times\langle t\rangle.
\end{equation}
The operators~\eqref{eq:GonV} and~\eqref{eq:GonO}, with the representation $r$ above, give a faithful unitary action of $G_{512}$ on $\mathcal N(L)$ by the same calculation~\eqref{eq:QR-lift-cocycle}--\eqref{eq:QR-lift-relations}. The image $r((-I_8,I_8,I_8))$ is $z=-I_{512}$, so $\mathcal T$ and $D_z\mathcal T$ again reverse the currents independently on the two sides.

Since $K$ is perfect and $r$ is faithful, Theorem~\ref{thm:Narain-anomaly} applies with $L=E_8^{\otimes3}$. At $D=512$ the $\mathcal T$-twisted spectrum~\eqref{eq:T-spectrum_Narain} becomes
\begin{equation}
    h=\frac{\lambda^2}{4}+N_L,\qquad
    \bar h=32+N_R^{\mathrm{tw}},
    \qquad \lambda\in L,\quad
    N_L\in\mathbb Z_{\ge0},\quad
    N_R^{\mathrm{tw}}\in\tfrac12\mathbb Z_{\ge0}.
    \label{eq:T512-T-spectrum}
\end{equation}
Evenness of $L$ gives $2(h-\bar h)\in\mathbb Z$, so the restriction to $\langle t\rangle$ vanishes by level-matching. However, unlike the QR and $P_{48q}$ cases, the restrictions to $K$ and $\langle t\rangle$ do not yet establish anomaly cancellation here, since the mixed summand $H_2(K,\mathbb Z)\otimes\mathbb Z_2$ in the K\"unneth decomposition~\eqref{eq:Kunneth_Narain} of $H_3(G_{512},\mathbb Z)$ can lead to a mixed anomaly. In the proof of Theorem~\ref{thm:Narain-anomaly} this summand is controlled by the uniqueness of the $\mathcal T$-twisted ground state. For $L=E_8^{\otimes3}$ the trace~\eqref{eq:ZT1_Narain} reads
\begin{equation}
    Z_{\mathcal T,1}
    =q^{-64/3}\bar q^{32/3}
      \left(1+512q+512\bar q^{1/2}+\cdots\right),
    \label{eq:T512-T-trace}
\end{equation}
whose coefficients count states in $\mathcal H_{\mathcal T}$ before the centralizer projection. Since a state of weights $(h,\bar h)$ contributes $q^{h-64/3}\bar q^{\bar h-64/3}$, the leading term in~\eqref{eq:T512-T-trace} is a unique ground state of weights $(0,32)$, on which the projective action of $K$ is trivial. Consequently $[\alpha]=0$ in $H^3(G_{512},U(1))$, so the full action can be gauged.

\subsection{\label{subsec:T512_Spectrum}The spectrum below and at dimension four}

In the untwisted sector, nonzero momentum--winding charges have $\Delta\ge\min(L)/2=4$. The eight-dimensional real representation $E_8\otimes_{\mathbb Z} \mathbb R$ of $H$ is absolutely irreducible~\cite{Atlas1985,Lubeck2021}, thus its tensor cube is absolutely irreducible for $H^{\times3}$. Hence it is an irreducible representation of $K$ and we have
\begin{equation}
    \operatorname{End}_K(L\otimes_{\mathbb Z}\mathbb R)=\mathbb R I_{512}.
    \label{eq:T512-commutant}
\end{equation}
By~\eqref{eq:GonO}, with $r(z)=-I_{512}$, the elements $\mathcal T$ and $D_z\mathcal T$ act on a neutral oscillator state by $(-1)^{s_R}$ and $(-1)^{s_L}$, respectively, where $s_L$ and $s_R$ are the numbers of left- and right-moving creation operators. Hence invariance under $\mathcal T$ and $D_z\mathcal T$ requires an even number of oscillators in each chiral sector. As in subsection~\ref{subsec:QR_Spectrum}, the only remaining neutral states below dimension four are the chiral bilinears at weights two and three. Then applying~\eqref{eq:T512-commutant} restricts their invariant coefficient matrices to be proportional to the identity, so they are vacuum descendants. In particular, there is no invariant antisymmetric matrix producing a weight-three primary, unlike the QR and $P_{48q}$ cases. Thus
\begin{equation}
    \left(\mathcal H_e^{G_{512}}\right)_{\Delta<4}=\operatorname{span}_{\mathbb C} \{\ket{0},L_{-2}\ket{0},\bar L_{-2}\ket{0}, L_{-3}\ket{0},\bar L_{-3}\ket{0}\}.
\end{equation}
Write $L_{\mathbb C}:=L\otimes_{\mathbb Z}\mathbb C$ for the complexified current space of the corresponding lattice in each case. It is worth emphasizing that the distinction is that in the QR case in subsection~\ref{subsec:QR_Spectrum} we have $r_{\mathbb C}=U\oplus \overline{U}$ hence $\operatorname{End}_K(r_{\mathbb C})=\mathbb C I\oplus \mathbb C W$, therefore $\operatorname{End}_K(L_{\mathbb C})\cap\mathfrak{so}(L_{\mathbb C})=\mathbb C W$. On the other hand, in this case we have irreducible $r_{\mathbb C}$ hence $\operatorname{End}_K(r_{\mathbb C})=\mathbb C I$, which gives $\operatorname{End}_K(L_{\mathbb C})\cap\mathfrak{so}(L_{\mathbb C})=\{0\}$.

For geometric twists $D_k$ with $k\ne1$, we bound the oscillator energy without listing conjugacy classes. Since $H=W(E_8)'\subset\mathrm{SO}(8)$, the nonreal eigenvalues of $h\in H$ occur in conjugate pairs with equal multiplicities. The multiplicities of $1$ and $-1$ are even, and multiplicity eight would force $h=I_8$ or $h=-I_8$. Hence every eigenvalue of a nonscalar element has multiplicity at most six. If one tensor factor of $H^{\times3}$ is nonscalar, the fixed space of $r(k)$ consequently has dimension at most $6\cdot64=384$, namely
\begin{equation}
	\operatorname{rank}(1-r(k))\ge128.
	\label{eq:nontrivial_rank_bound}
\end{equation}

Since $r(k)$ preserves $L$, its matrix in a lattice basis has integer entries, so its characteristic polynomial has integer coefficients, namely $\chi_{r(k)}(x)=\det(xI-r(k))\in\mathbb Z[x]$. The roots of $\chi_{r(k)}(x)$ are roots of unity because $r(k)$ has finite order. Since the cyclotomic polynomial $\Phi_n(x)=\prod_{\substack{1\le j\le n\\(j,n)=1}} \left(x-e^{2\pi i j/n}\right)$ is irreducible over $\mathbb Q$ and is the minimal polynomial of every primitive $n$-th root of unity~\cite[Thm.~5.10]{Milne2003}, $\chi_{r(k)}(x)$ must factor into powers of $\Phi_n(x)$'s. Consequently, whenever a primitive $n$-th root occurs as a root of $\chi_{r(k)}(x)$, all primitive $n$-th roots must occur with the same multiplicity. We call one copy of this complete set a cyclotomic block of order $n$. Each nontrivial block has $n>1$ and contains $\varphi(n)$ eigenvalues. Thus, for a nontrivial block, the average of $\theta(1-\theta)$ over its eigenangles satisfies
\begin{equation}
    \frac{\sum_{\theta\in\text{block}}\theta(1-\theta)} {\text{block size}} =\frac1{\varphi(n)}\sum_{\substack{1\le j<n\\(j,n)=1}}\frac jn\left(1-\frac jn\right)=\frac16\left(1-\frac{\prod_{\ell\mid n,\ \ell\ {\rm prime}}(-\ell)}{n^2}\right)
    \ge\frac5{36},
    \label{eq:cyclotomic_sum}
\end{equation}
where $\varphi$ is Euler's totient function. To derive the equality in~\eqref{eq:cyclotomic_sum}, we note that the roots of $x^t-1$ are precisely the eigenvalues of a positive cycle of length $t$ in subsection~\ref{subsec:QR_Spectrum}, and they consist of exactly one cyclotomic block of each order $d\mid t$. Writing $S(n)=\sum_{\theta\in\text{block}}\theta(1-\theta)$ for a block of order $n$, the value of $\rho_+(t)$ computed there gives
\begin{equation}
    \sum_{d\mid t}S(d)=4\rho_+(t)=\frac16\left(t-\frac1t\right).
\end{equation}
Letting $\mu$ denote the M\"obius function, M\"obius inversion~\cite{Apostol1976} then gives
\begin{equation}
    S(n)=\frac16\sum_{t\mid n}\mu\!\left(\frac nt\right)\left(t-\frac1t\right)=\frac{\varphi(n)}6-\frac1{6n}\prod_{\ell\mid n,\ \ell\ {\rm prime}}(1-\ell),
\end{equation}
where for the last equality we use the standard divisor-sum formulas $\sum_{t\mid n}\mu(n/t)\,t=\varphi(n)$ and $\sum_{t\mid n}\mu(n/t)\,t^{-1}=n^{-1}\sum_{d\mid n}\mu(d)\,d$~\cite[Thms.~2.3 and 2.18]{Apostol1976}. Equivalently, $\Phi_n(x)=\prod_{t\mid n}(x^t-1)^{\mu(n/t)}$ has cycle shape $\prod_{t\mid n}t^{\mu(n/t)}$, and $S(n)/4$ is its vacuum anomaly $\frac1{24}\sum_tb_t\left(t-\frac1t\right)$ in the notation of~\cite[Sec.~3.1]{MollerScheithauer2019}. Dividing $S(n)$ by Euler's product $\varphi(n)=n\prod_{\ell\mid n,\ \ell\ {\rm prime}}(1-1/\ell)$~\cite[Thm.~2.4]{Apostol1976} we obtain the equality in~\eqref{eq:cyclotomic_sum}. For the lower bound, we note that the product $\prod_{\ell\mid n,\ \ell\ {\rm prime}}(-\ell)$ is negative when $n$ has an odd number of distinct prime divisors. Hence to achieve the lower bound we must require $n$ have an even number of prime divisors, thus $n\ge6$. Therefore we have $\frac{\prod_{\ell\mid n,\ \ell\ {\rm prime}}\ell}{n^2}\le1/n\le1/6$, which gives the bound $5/36$, with equality at $n=6$.

Since $r(k)$ is diagonalizable, $\operatorname{rank}(1-r(k))$ equals the total number of eigenvalues different from one, counted with multiplicity, and hence the total size of all nontrivial blocks. Therefore, plugging~\eqref{eq:nontrivial_rank_bound} and~\eqref{eq:cyclotomic_sum} into~\eqref{eq:lattice-twist-energy} implies
\begin{equation}
	\rho_{\mathrm{osc}}(r(k))\ge\frac{5}{4\times36}\operatorname{rank}(1-r(k))\ge \frac{40}{9}
\end{equation}
for a nonscalar $r(k)$. While if every factor of $H^{\times3}$ is scalar, the only nonidentity image is $z=-I_{512}$, with $\rho_{\mathrm{osc}}(z)=32$. Combining the two cases, and using the same eigenangle inequality as in~\eqref{eq:P48q-twisted-bound} for twists containing $\mathcal T$, gives
\begin{equation}
    \Delta\ge
    \begin{cases}
        80/9,&\text{in }\mathcal H_{D_k},\quad k\ne1,\\
        32,&\text{in }\mathcal H_{D_k\mathcal T}.
    \end{cases}
    \label{eq:T512-twisted-bound}
\end{equation}
Adding nonnegative zero--mode momentum--winding contributions or performing centralizer projection preserves these lower bounds. Thus every primary at dimension four must be in the untwisted sector. For any norm-eight vector $\mu\in L$, the charge-orbit sum~\eqref{eq:OrbitSum_QR} is a nonzero invariant primary of weights $(2,2)$, which establishes the exact gap at four.

\begin{theorem}[Gap of the rank-512 tensor-lattice orbifold]
\label{thm:T512-gap}
Let $L=E_8^{\otimes3}$ and let $K$ act as in~\eqref{eq:T512-current-action}. The orbifold
\begin{equation}
    \mathcal C_{T_{512}}=\mathcal N(L)/(K\times\langle t\rangle)
\end{equation}
with the lifted action defined by~\eqref{eq:GonV} and~\eqref{eq:GonO} is a compact unitary bosonic full CFT with a unique vacuum, $(c_L,c_R)=(512,512)$ and $\Delta_{\mathrm{gap}}=4$. The theory is nonfactorizing.
\end{theorem}

Unitarity and compactness follow from the same finite gauging construction as for the QR family, while~\eqref{eq:T512-twisted-bound} excludes additional vacua. The nonfactorization proof in Theorem~\ref{thm:QR-gap} also applies, since the invariant primary $Q_{2,2}$ has the positive norm~\eqref{eq:mixed-primary-norm} at $D=512$, whereas there is no primary of weight $(2,0)$ or $(0,2)$.

\section{\label{sec:full-BW}Barnes--Wall Orbifolds with Virasoro Primary Gap Four}

In this section we construct a family of 2d full CFTs with Virasoro primary gap $\Delta_{\mathrm{gap}}=4$. Such CFTs are obtained from orbifolding parent CFTs associated with the Barnes-Wall lattices $BW(m)$ by the groups $\widetilde G=E(m)\times_A E(m)$ that correlate two extraspecial actions through $A=E(m)/Z(E(m))$. In subsection~\ref{subsec:BW_Quotient}, we show that lifted $\widetilde G$-action on the parent CFT is anomaly-free. In subsection~\ref{subsec:BW_Spectrum} we determine the low-dimension spectrum and show that all theories $\mathcal C_{BW(m)}/\widetilde G$ in this family have Virasoro primary gap four.

\subsection{\label{subsec:BW_Quotient}Anomaly-free \texorpdfstring{$\widetilde G$}{G-tilde} action on \texorpdfstring{$\mathcal C_{BW(m)}$}{the Barnes--Wall parent}}

In this subsection we show that there exists an anomaly-free $\widetilde G$ action on the CFT $\mathcal C_{BW(m)}$ associated with the Barnes--Wall lattice $BW(m)$ where $E(m)$ is an extraspecial 2-group and $\widetilde G=E(m)\times_A E(m)$. The main result of this subsection is summarized as the following theorem
\begin{theorem}[Anomaly cancellation for Barnes--Wall orbifolds]
\label{thm:BW-anomaly}
Let $m\ge7$ be odd, let $V=V_{BW(m)}$ where $BW(m)$ is the even unimodular Barnes--Wall lattice of rank $2^m$, and let $G=E(m)\cong 2_+^{1+2m}$ be the extraspecial 2-group acting on $V$ through the faithful lift of~\cite[Theorem~3.5]{KRR2026}. Write $A=G/Z(G)$ and let $\pi:G\to A$ be the quotient map. Then the chiral anomaly is inflated from $A$,
\begin{equation}
    [\alpha_V]=\pi^*[\nu]\quad\text{for some }[\nu]\in H^3(A,U(1)).
\end{equation}
Consequently, the action of
\begin{equation}
    \widetilde G=G\times_A G = \{(g_L,g_R)\in G\times G:\pi(g_L)=\pi(g_R)\}
    \label{eq:BW-fiber-product}
\end{equation}
on $V\otimes\overline V$ is faithful and anomaly-free,
\begin{equation}
    [\alpha_{\mathrm{full}}] = 0\in H^3(\widetilde G,U(1)),
\end{equation}
and therefore admits consistent finite gauging by $\widetilde G$.
\end{theorem}

The fiber product is chosen so that the two chiral anomalies cancel if they come from the same class on $A$. To see this, consider a split parent $V\otimes\overline{V'}$ with symmetry groups $G_L,G_R$ and quotient maps $\pi_L:G_L\to A$ and $\pi_R:G_R\to A$. The group
\begin{equation}
    \widetilde G=G_L\times_A G_R
    =\{(g_L,g_R)\mid \pi_L(g_L)=\pi_R(g_R)\}
    \label{eq:GtildeAction}
\end{equation}
acts by
\begin{equation}
    (g_L,g_R)(v_L\otimes\overline{v_R})
    =g_Lv_L\otimes\overline{g_Rv_R}.
    \label{eq:split-action}
\end{equation}
Write $\operatorname{pr}_L,\operatorname{pr}_R$ for the projections onto the two groups. The full anomaly is $\operatorname{pr}_L^*[\alpha_L]-\operatorname{pr}_R^*[\alpha_R]$. If $[\alpha_L]=\pi_L^*[\nu]$ and $[\alpha_R]=\pi_R^*[\nu]$ for the same class $[\nu]\in H^3(A,U(1))$, the two terms agree because $\pi_L\circ \operatorname{pr}_L=\pi_R\circ \operatorname{pr}_R$.

For the Barnes--Wall parent the two chiral factors carry the same $G$-action, so it remains to show that its chiral anomaly $[\alpha]$ is inflated from $A$. For the group-theoretic argument we take $G=E(m)$ with $m\ge3$ and write
\begin{equation}
    1\longrightarrow\langle z\rangle\longrightarrow G
    \xrightarrow{\pi}A\longrightarrow1,
    \qquad A=G/\langle z\rangle\cong(\mathbb Z_2)^{2m},
    \label{eq:SES_NGA}
\end{equation}
where $z$ is the central involution. The extension defines a quadratic form $Q$ on $A$ by $g^2=z^{Q(\pi(g))}$~\cite[Sec.~3.3]{JFT2020}.
We will show that for any $g$ with $g^2=z$, the class $[\alpha]$ is inflated precisely when its restriction to $\langle g\rangle$ vanishes.
This reduces the anomaly calculation to level matching in a single order-four twisted sector.

For extraspecial 2-groups the inflated classes are characterized by their order. Indeed, for $m\ge3$ we have~\cite[Sec.~3.3]{JFT2020}
\begin{equation}
	H^3(G,U(1))\cong(\mathbb Z_2)^{n}\oplus\mathbb Z_4
	\label{eq:BW-H3}
\end{equation}
for some integer $n\ge0$, and since $G\not\cong D_8$, the classes of order at most two are exactly the inflated ones~\cite[Lemma~4.2]{GoffMasonNg2007},
\begin{equation}
	\operatorname{im}\pi^*=\left\{x\in H^3(G,U(1)):2x=0\right\}.
	\label{eq:BW-inflation-2torsion}
\end{equation}
By~\eqref{eq:BW-H3}, the subgroup $\operatorname{im}\pi^*$ therefore has index two in $H^3(G,U(1))$, so we only need to distinguish its two cosets.\footnote{The assumption $m\ge3$ is necessary. For $m=2$ the group $H^3(G,U(1))$ contains elements of order eight~\cite[Sec.~3.3]{JFT2020}, so \eqref{eq:BW-H3} fails, and so do the statements of~\cite[Lemmas~4.1(a) and~4.5]{GoffMasonNg2007} in that case.} For this we choose an arbitrary $a\in A$ with $Q(a)=1$ and set $B=\langle a\rangle$, so that $Q|_B\ne0$, and such an $a$ must exist because $Q$ has nondegenerate polarization. Thus any lift $g\in G$ with $\pi(g)=a$ satisfies $g^2=z$, hence
\begin{equation}
    C=\pi^{-1}(B)=\{1,z,g,gz\}=\langle g\rangle\cong\mathbb Z_4.
    \label{eq:DefineC}
\end{equation}
We now show that restriction to $C$ distinguishes the two cosets of $\operatorname{im}\pi^*$.

Let $X=\operatorname{im}\pi^*$ and $Y=\ker\operatorname{res}_C^G$ inside $H^3(G,U(1))$. Since the connecting isomorphism $\delta: H^3(- ,U(1))\cong H^4(-, \mathbb Z)$ for finite groups is natural with respect to inflation and restriction~\cite{Weibel1994}, we may compare $X$ and $Y$ in integral cohomology. We consider the quotient $q:C\to B$ induced by $\pi$. The cyclic cohomology rings are $H^*(C,\mathbb Z)=\mathbb Z[u]/(4u)$ and $H^*(B,\mathbb Z)=\mathbb Z[v]/(2v)$, with $|u|=|v|=2$~\cite{Weibel1994}. We choose $u$ and $v$ to correspond to the characters $\chi_C(g) = i$ and $\chi_B(a) = -1$. Pullback along $q$ then gives $q^*\chi_B = \chi_C^2$ which in terms of cohomology classes gives
\begin{equation}
    q^*v=2u,\qquad q^*(v^2)=4u^2=0.
\end{equation}
Thus $q^*$ vanishes in degree four, namely $q^*:H^4(B,\mathbb Z)\longrightarrow H^4(C,\mathbb Z)$ is the zero map. The commutativity of the following diagram
\begin{equation}
	\begin{array}{ccc}
		H^4(A,\mathbb Z)&\xrightarrow{\ \pi^*\ }&H^4(G,\mathbb Z)\\
		{\scriptstyle\operatorname{res}_B^A}\downarrow &&\downarrow{\scriptstyle\operatorname{res}_C^G}\\
		H^4(B,\mathbb Z)&\xrightarrow{\ q^*\ }&H^4(C,\mathbb Z).
	\end{array}
\end{equation}
then gives
\begin{equation}
    \operatorname{res}_C^G\circ\pi^*
    =q^*\circ\operatorname{res}_B^A=0.
\end{equation}
Namely, $X\subseteq Y$.

On the other hand, for $m\ge3$ we have $|\operatorname{im}\operatorname{res}_C^G|=2$~\cite[Lemma~4.5]{GoffMasonNg2007}, so $Y$ also has index two in $H^3(G,U(1))$. Therefore, since $X\subseteq Y$, we have
\begin{equation}
    [H^3(G,U(1)):X] = [H^3(G,U(1)):Y][Y:X] \Rightarrow [Y:X] = 1
    \label{eq:BW-order-four-inflation}
\end{equation}
since $[H^3(G,U(1)):X] = [H^3(G,U(1)):Y] = 2$. Therefore, we have $\operatorname{im}\pi^* = \ker\operatorname{res}_C^G$. Consequently, restricting $G$ to any $C$ in the form~\eqref{eq:DefineC} determines whether an anomaly class is inflated from $A$, namely we have
\begin{equation}
    [\alpha]=\pi^*[\nu]\ \ \text{for some }[\nu]\in H^3(A,U(1))
    \quad\Longleftrightarrow\quad
    \operatorname{res}_C^G[\alpha]=0.
    \label{eq:BW-inflation-criterion}
\end{equation}

To test if $\operatorname{res}_C^G[\alpha]=0$ we apply level-matching~\eqref{eq:cyclic-type} again. Let $\rho_g$ denote the lowest conformal weight of the $g$-twisted sector. Recall from~\eqref{eq:cyclic-type} that the restricted anomaly is determined by the type
\begin{equation}
	t_g=4^2\rho_g\pmod{4}\in\mathbb Z_4
\end{equation}
where we have focused on $N_g = 4$ case. In particular, we have
\begin{equation}
	\operatorname{res}_C^G[\alpha]=0 \quad\Longleftrightarrow\quad t_g=0 \quad\Longleftrightarrow\quad 4\rho_g\in\mathbb Z.
	\label{eq:tg_LevelMatching}
\end{equation}

We now restrict ourselves to Barnes-Wall lattices with $D=2^m$ for the lattice rank with odd $m\ge7$~\cite{BarnesWall1961} and let $G_L = G_R = G = E(m)$ in~\eqref{eq:GtildeAction}~\cite{KRR2026}. The central element $z$ acts as $-I_D$ on the currents~\cite{KRR2026}. Since our chosen element satisfies $g^2=z$, its action satisfies
\begin{equation}
	R(g)^2=-I_D.
\end{equation}
Hence its eigenvalues are $i$ and $-i$, each with multiplicity $D/2$. Since the twist removes all bosonic momentum zero modes, the ground-state conformal weight comes entirely from the shifted oscillator vacuum energy. Therefore, we have
\begin{equation}
	\rho_g = \rho_{\mathrm {osc}}(R(g)) = \frac14 \frac D2\left( \left(\frac14\right)\left(\frac34\right) + \left(\frac34\right)\left(\frac14\right) \right) = \frac{3D}{64}.
\end{equation}
Consequently, we have $4\rho_g = \frac{3D}{16} = 3\cdot2^{m-4}\in\mathbb Z$ since $m \ge 7$. Thus by~\eqref{eq:tg_LevelMatching}, we have $\operatorname{res}_C^G[\alpha]=0$. Therefore, by~\eqref{eq:BW-inflation-criterion}, there must exist $[\nu]\in H^3(A,U(1))$ such that $[\alpha]=\pi^*[\nu]$. Namely, the chiral anomaly comes from $A$. For our identical left and right VOA constructions, the same class $[\nu]$ occurs on both sides. Therefore, on the total group
\begin{equation}
	\widetilde G=G\times_A G,
\end{equation}
the defining relation $\pi\circ \operatorname{pr}_L=\pi\circ \operatorname{pr}_R$ leads to
\begin{equation}
	[\alpha_{\mathrm{full}}] = \operatorname{pr}_L^*[\alpha]-\operatorname{pr}_R^*[\alpha] = (\pi\circ \operatorname{pr}_L)^*[\nu]-(\pi\circ \operatorname{pr}_R)^*[\nu] = 0.
\end{equation}
Therefore, the lift of~\cite{KRR2026} induces a faithful unitary action of $E(m)\times_A E(m)$ on $V\otimes\overline V$, and we have proved that this action is anomaly-free.

\subsection{\label{subsec:BW_Spectrum}The Spectrum of \texorpdfstring{$\mathcal C_{BW(m)}/\widetilde G$}{the Barnes--Wall Orbifold}}

We have shown that the parent theory of the split construction is $\mathcal C_{BW(m)} = V\otimes\overline V$ with $V=V_{BW(m)}$, whose momentum lattice is $\Gamma_{\mathrm{split}}(L)=L\oplus L(-1)$ of~\eqref{eq:narain-parent-points}. Its left and right charges are therefore independent,
\begin{equation}
	h=\frac{p_L^2}{2}+N_L, \qquad \bar h=\frac{p_R^2}{2}+N_R, \qquad p_L,p_R\in L,
\end{equation}
in contrast with the correlated momentum and winding labels of section~\ref{sec:CFTs}. The Barnes--Wall lattice of rank $D=2^m$ has minimum squared norm $2^{(m-1)/2}$~\cite{BarnesWall1961,KRR2026}, so every state with $(p_L,p_R)\ne(0,0)$ obeys
\begin{equation}
	\Delta\ge\frac{\min(L)}{2}=2^{(m-3)/2}\ge4
	\label{eq:BW_ChargeBound}
\end{equation}
for odd $m\ge7$. Thus asking $\Delta<4$ forces $p_L=p_R=0$, and the remaining states consist entirely of oscillator excitations.

Both central elements $(z,1)$ and $(1,z)$ lie in $\widetilde G$. Each acts as $-I_D$ on its own chiral current space and as the identity on the other. They therefore act on a neutral state by $(-1)^{n_L}$ and $(-1)^{n_R}$, where $n_L$ and $n_R$ count the left- and right-moving creation operators, so an invariant neutral state carries an even number of them in each chiral sector. Below $\Delta=4$ this excludes every left-right mixed state, together with $a^i_{-3}\ket{0}$ and $a^i_{-1}a^j_{-1}a^k_{-1}\ket{0}$, and leaves only the chiral bilinears
\begin{equation}
	\sum_{i,j}A_{ij}\,a^i_{-1}a^j_{-1}\ket{0}, \qquad \sum_{i,j}A_{ij}\,a^i_{-2}a^j_{-1}\ket{0}
\end{equation}
of weights two and three, together with their right-moving counterparts. The real $E(m)$-module $L\otimes_{\mathbb Z}\mathbb R$ is absolutely irreducible~\cite{CalderbankCameronKantorSeidel1997,KRR2026}, hence by Schur's lemma its commutant is $\mathbb R I_D$. Therefore the only invariant matrix $A$, up to scale, is $A=\delta$. At weight two this leaves $\sum_i a^i_{-1}a^i_{-1}\ket{0}=2L_{-2}\ket{0}$ and at weight three it leaves $\sum_i a^i_{-2}a^i_{-1}\ket{0}=L_{-3}\ket{0}$, both of which are vacuum descendants. In particular there is no invariant antisymmetric matrix here, so, unlike the QR family of section~\ref{sec:CFTs}, the weight-three bilinears contain no primary. Therefore
\begin{equation}
	\left(\mathcal H_e^{\widetilde G}\right)_{\Delta<4} = \operatorname{span}_{\mathbb C}\left\{ \ket{0},\ L_{-2}\ket{0},\ \overline L_{-2}\ket{0},\ L_{-3}\ket{0},\ \overline L_{-3}\ket{0} \right\},
\end{equation}
and every state below dimension four lies in the vacuum Virasoro module.

For the twisted sectors, recall that every element of $\widetilde G$ has the form $(g,z^\varepsilon g)$ with $g\in E(m)$ and $\varepsilon\in\{0,1\}$. Apart from the identity, $E(m)$ contains the central element $z$, with
$R(z)=-I_D$, the noncentral involutions, with $R(g)^2=I_D$, and the
elements of order four, with $R(g)^2=-I_D$. For any noncentral $g$ the commutator map $h\mapsto[h,g]=hgh^{-1}g^{-1}$ onto $\langle z\rangle$ is surjective, so $g$ and $zg$ are conjugate due to $[h,g] = z$ thus
\begin{equation}
	\operatorname{tr}R(g)=\operatorname{tr}R(zg)=-\operatorname{tr}R(g)=0.
\end{equation}
Hence the two eigenvalues $\pm1$ of $R(g)$ for noncentral involution, respectively $\pm i$ of $R(g)$ for $g$ with order $4$, occur with equal multiplicity $D/2$, and~\eqref{eq:lattice-twist-energy} gives
\begin{equation}
	\rho_{\mathrm{osc}}(R(z))=\frac D{16}, \qquad
	\rho_{\mathrm{osc}}(R(g)) = \begin{cases}
		\frac{D}{32}, & g\text{ a noncentral involution},\\
		\frac{3D}{64}, & g\text{ of order four}.
	\end{cases}
	\label{eq:BW-chiral-twist-energies}
\end{equation}
Since $z$ is central, $zg$ has the same type as $g$, namely either both are noncentral involution or both are of order four. Together with~\eqref{eq:GtildeAction}  which ensures $g_L = z^\varepsilon g_R$, the paired bound $\Delta\ge\rho_{\mathrm{osc}}(R(g_L))+\rho_{\mathrm{osc}}(R(g_R))$ equals $D/16$ for $(z,1)$ and $(1,z)$, $D/8$ for $(z,z)$, $D/16$ for a pair of noncentral involutions and $3D/32$ for a pair of order four. Consequently
\begin{equation}
	\Delta\ge\frac D{16}=2^{m-4}\ge8>4 \qquad\text{for every state of }\mathcal H_g\text{ with }[g]\ne[e],
	\label{eq:BW_TwistedBound}
\end{equation}
where for the numerical bound we used $m\ge7$. Since the projections only remove states, the same bound holds after projection, and together with the untwisted analysis we obtain $\Delta_{\mathrm{gap}}\ge4$.

It remains to exhibit a primary at $\Delta=4$. Set
\begin{equation}
	A_4=\left(\sum_i a^i_{-1}a^i_{-1}\right)^2\ket{0}, \qquad B_4=\sum_i a^i_{-2}a^i_{-2}\ket{0}, \qquad C_4=\sum_i a^i_{-3}a^i_{-1}\ket{0},
\end{equation}
and write $S=\sum_i a^i_{-1}a^i_{-1}\ket{0}$ and $U=\sum_i a^i_{-2}a^i_{-1}\ket{0}$. Using $[L_n,a^i_k]=-k\,a^i_{n+k}$ together with $a^i_k\ket{0}=L_k\ket{0}=0$ for $k\ge0$, we find
\begin{equation}
	\begin{array}{c|ccc}
		& A_4 & B_4 & C_4\\ \hline
		L_1 & 0 & 4U & 3U\\
		L_2 & 2(D+2)S & 0 & 3S\\
		L_3 & 0 & 0 & 0\\
		L_4 & 0 & 4D\ket{0} & 3D\ket{0}
	\end{array}
\end{equation}
while every $L_n$ with $n\ge5$ annihilates a state of weight four. The combination
\begin{equation}
	P_D = 6A_4+3(D+2)B_4-4(D+2)C_4
	\label{eq:BW_P4}
\end{equation}
is therefore a Virasoro primary of weights $(4,0)$, and it is nonzero because
\begin{equation}
	\|P_D\|^2 = 24D(D+2)(5D+22)>0.
\end{equation}
Its three terms carry four, two and two left-moving creation operators and none on the right, so both central signs act trivially, and it is built from $\delta$ alone, hence invariant under $E(m)$. The right-moving primary $\overline P_D$ at $(0,4)$ is obtained by replacing $a^i_{-n}$ with $\bar a^i_{-n}$. This establishes the exact gap. It does not enumerate all weight-four primaries, which can also receive contributions from invariant quartic tensors and, at $m=7$, lattice vectors. We summarize the result as follows.

\begin{theorem}[Gap of the Barnes--Wall orbifolds]
\label{thm:BW-gap}
Let $m\ge7$ be odd, let $\mathcal C_{BW(m)}=V_{BW(m)}\otimes\overline{V_{BW(m)}}$, and let $\widetilde G=E(m)\times_A E(m)$ act as in Theorem~\ref{thm:BW-anomaly}. Then the orbifold $\mathcal C_{BW(m)}/\widetilde G$ is a compact unitary bosonic full CFT with a unique vacuum and $(c_L,c_R)=(2^m,2^m)$, and it has
\begin{equation}
	\Delta_{\mathrm{gap}}\!\left(\mathcal C_{BW(m)}/\widetilde G\right)=4.
\end{equation}
Moreover $\mathcal C_{BW(m)}/\widetilde G$ is nonfactorizing.
\end{theorem}

Unitarity, compactness and uniqueness of the vacuum follow exactly as in section~\ref{sec:CFTs}, with~\eqref{eq:BW_TwistedBound} in place of~\eqref{eq:TwistedBound_QR}. The nonfactorization argument applies verbatim as well. The orbifold has no $(2,0)$ primary, so a splitting into a holomorphic and an antiholomorphic factor would force $P_{2,2}(\mathcal H_{\mathrm{orb}})=0$, whereas $Q_{2,2}$ is again a nonzero $\widetilde G$-invariant $(2,2)$ primary, carrying two creation operators in each chiral sector and built from $\delta$ alone. The independent central signs act trivially because each side contains two oscillators. Here the $(2,2)$ primaries can in fact be counted exactly. Using the standard matrix realization of the extraspecial group~\cite{CalderbankCameronKantorSeidel1997}, the conjugation action of $A=E(m)/Z(E(m))$ on $\operatorname{End}(W)$, with $W=L\otimes_{\mathbb Z}\mathbb R$, is multiplicity-free. The one-dimensional character line labelled by $a\in A$ is spanned by a symmetric matrix precisely when $Q(a)=0$. Thus removing the $a=0$ line, which is spanned by the identity matrix, leads to
\begin{equation}
    \operatorname{Sym}_0^2(W)
    \cong
    \bigoplus_{\substack{a\ne0\\ Q(a)=0}}\chi_a .
\end{equation}
For the full theory, the two factors of $\widetilde G=E(m)\times_A E(m)$ have the same image in $A$. Hence $\chi_a\otimes\chi_b$ is $\widetilde G$-invariant precisely when $a=b$, so each nonzero singular vector contributes one $(2,2)$ primary. Since $Q$ is of plus type, there are $2^{2m-1}-2^{m-1}$ nonsingular vectors~\cite[Theorem~6.14, proof~(i)]{HallShpectorov2021}. Subtracting these and the zero vector from all $2^{2m}$ vectors gives
\begin{equation}
    2^{2m-1}+2^{m-1}-1
    =
    \frac{D(D+1)}{2}-1 = \dim \operatorname{Sym}_0^2(W) .
\end{equation}
Thus there are $8255$ such primaries for $m=7$ and $131327$ for $m=9$, etc.

\section{\label{sec:conclusion}Conclusion and Outlook}

We have constructed compact unitary bosonic full CFTs in two dimensions with a unique vacuum and exact all-spin Virasoro-primary gaps three and four, resolving the existence problem posed in~\cite{KRR2026}.
The quadratic-residue family has gap three at $c_L=c_R=p+1$ for every prime $p\ge71$ with $p\equiv23\pmod{24}$, and the Barnes--Wall family has gap four at $c_L=c_R=2^m$ for every odd $m\ge7$.
The two additional examples have gap three at $c_L=c_R=48$ and gap four at $c_L=c_R=512$. All these theories are nonfactorizing.

The constructions combine a large lattice minimum with a common orthogonal action on the left- and right-moving currents and independent chiral sign reversals.
The lattice minimum excludes light charged untwisted states, while the sign reversals remove the dimension-two current bilinear preserved by identical left and right actions.
The remaining low-weight invariant tensors and the bounds on twisted-sector ground-state energies determine the exact gaps.
For the Narain parents, consistency of the gauging follows from the general anomaly-cancellation result of Theorem~\ref{thm:Narain-anomaly}.
For the split Barnes--Wall parents, the two chiral anomalies are inflated from the same class on a common quotient and cancel under correlated gauging, without requiring either chiral anomaly to vanish separately.

The weight-four chiral primary $P_D$ in~\eqref{eq:BW_P4} also identifies a limitation of this approach.
It is nonzero and built entirely from contractions with the Euclidean metric, so it is invariant under the full orthogonal group $O(D)$.
Consequently, it survives in the untwisted sector of any anomaly-free finite gauging of a lattice CFT whose symmetry action preserves the Heisenberg current space on each chiral side and acts there by orthogonal transformations.
Such orbifolds therefore satisfy $\Delta_{\mathrm{gap}}\le4$, even when the left- and right-moving current actions are chosen independently.
Our gap-four examples saturate this bound.

A direct next step is to apply these constructions and anomaly-cancellation arguments to other lattices and finite symmetry groups, with particular interest in realizing gap four at smaller central charges.
Obtaining an all-spin Virasoro-primary gap greater than four requires a construction that avoids the invariant primary described above.
A further question is whether explicit families of full CFTs can have gaps that grow parametrically with their central charges.
Such families would provide benchmarks for the modular bootstrap~\cite{Hellerman2011,CollierLinYin2018} and help clarify the spectral requirements for putative CFT duals of pure gravity in $\mathrm{AdS}_3$~\cite{Witten2007,HKS2014}.

\begin{appendices}
\renewcommand*{\theHequation}{\thesection.\arabic{equation}}

\section{\label{app:Narain-anomaly}Proof of anomaly cancellation in Narain examples}

In this appendix we prove Theorem~\ref{thm:Narain-anomaly}. Throughout, $L$ is a positive-definite even unimodular lattice of rank $D$, $K$ is a finite perfect group with a faithful representation $r:K\to O(L)$, and $G=K\times\langle t\rangle$ acts on $\mathcal N(L)$ by~\eqref{eq:GonV} and~\eqref{eq:GonO}. Recall that the rank of an even unimodular lattice satisfies $D\in8\mathbb Z$~\cite[Ch.~V]{Serre1973}. We first decompose the group in which the anomaly lives, and then show that $[\alpha]$ pairs trivially with each summand.

\paragraph{Decomposition of the anomaly}
Since $K$ is perfect, we have $H_1(K,\mathbb Z)=0$. Together with $H_1(\mathbb Z_2,\mathbb Z)=\mathbb Z_2$, $H_2(\mathbb Z_2,\mathbb Z)=0$ and $H_3(\mathbb Z_2,\mathbb Z)=\mathbb Z_2$~\cite{Brown1982}, the K\"unneth theorem gives
\begin{equation}
    H_3(G,\mathbb Z)\cong
    H_3(K,\mathbb Z)\oplus H_3(\mathbb Z_2,\mathbb Z)
    \oplus\bigl(H_2(K,\mathbb Z)\otimes\mathbb Z_2\bigr)
    \label{eq:Kunneth_Narain}
\end{equation}
where the terms involving $H_1(K,\mathbb Z)$, including the torsion term $\operatorname{Tor}(H_1(K,\mathbb Z),\mathbb Z_2)$, vanish by perfectness of $K$. Since $U(1)$ is divisible, the universal coefficient theorem gives $H^3(G,U(1))\cong\operatorname{Hom}(H_3(G,\mathbb Z),U(1))$, so $[\alpha]$ is determined by its pairings with the three summands in~\eqref{eq:Kunneth_Narain}. Since $K$ and $\mathbb Z_2=\langle t\rangle$ enter $G$ through inclusion, the pairings with the first two summands are captured by the restrictions $\operatorname{res}^{G}_{K}[\alpha]$ and $\operatorname{res}^{G}_{\langle t\rangle}[\alpha]$. If $K$ is superperfect, as in the QR and $P_{48q}$ cases, then $H_2(K,\mathbb Z)=0$ and we simply have
\begin{equation}
	H^3(G,U(1))\cong H^3(K,U(1))\oplus H^3(\mathbb Z_2,U(1)),\qquad
	[\alpha]\longmapsto \left( \operatorname{res}^{G}_{K}[\alpha], \operatorname{res}^{G}_{\langle t\rangle}[\alpha] \right).
	\label{eq:H3Isom_QR}
\end{equation}
In general, however, the mixed summand $H_2(K,\mathbb Z)\otimes\mathbb Z_2$ can lead to a mixed anomaly and has to be treated separately.

\paragraph{Restriction to \texorpdfstring{$K$}{K}}
We first consider $\operatorname{res}^{G}_{K}[\alpha]$. The parent theory $\mathcal N(L)$ admits the standard Dirichlet boundary condition at the origin of the target torus, corresponding to a D0-brane localized there. To describe its boundary state, take the worldsheet boundary to be the unit circle $z=e^{i\sigma}$. The Dirichlet condition requires every point on this circle to map to the target-space origin,
\begin{equation}
	X(e^{i\sigma},e^{-i\sigma})=0\pmod{2\pi L}.
\end{equation}
Thus the tangential derivative vanishes along the boundary. Using the chiral mode expansions
\begin{equation}
	\partial X_L^i(z)=-i\sum_{n\in\mathbb Z}a_n^i z^{-n-1},
	\qquad
	\bar\partial X_R^i(\bar z)=-i\sum_{n\in\mathbb Z}\bar a_n^i\bar z^{-n-1},
\end{equation}
this condition becomes
\begin{equation}
	0=iz\,\partial X_L^i-i\bar z\,\bar\partial X_R^i
	=\sum_{n\in\mathbb Z}(a_n^i-\bar a_{-n}^i)e^{-in\sigma}.
\end{equation}
Consequently, the boundary state satisfies
\begin{equation}
	(a_n^i-\bar a_{-n}^i)\ket{D_0}=0
	\qquad\text{for all }n\in\mathbb Z.
\end{equation}
Since $a_0^i$ and $\bar a_0^i$ act on $\ket{m,w}$ with eigenvalues $p_L^i$ and $p_R^i$, respectively, the zero-mode condition requires $p_L=p_R$. By $p_L-p_R=\sqrt2\,w$, only sectors with $w=0$ contribute. The nonzero-mode conditions determine the positive sign in the oscillator exponential, while the Dirichlet condition localizes the zero mode at the target-space origin and its Fourier expansion gives a sum over all momentum states with equal coefficients, hence the zero-mode part is proportional to $\sum_{m\in L}\ket{m,0}$. The resulting boundary state is therefore
\begin{equation}
	\ket{D_0}
	=2^{-D/4}
	\exp\!\left(\sum_{n=1}^{\infty}\frac1n\sum_{i=1}^{D}a_{-n}^{i}\bar a_{-n}^{i}\right)
	\sum_{m\in L}\ket{m,0},
	\label{eq:D0-state}
\end{equation}
which generalizes the compact-boson Dirichlet state in~\cite[Eqs.~(82b)--(83b)]{DiGiulio2023}.

To check its normalization and simplicity, take $\tau\in i\mathbb R_{>0}$, $q=e^{2\pi i\tau}$, $\widetilde\tau=-1/\tau$, and $\widetilde q=e^{2\pi i\widetilde\tau}$. With $\Theta_L(\tau)=\sum_{\lambda\in L}e^{\pi i\tau\lambda^2}$, the annulus overlap is
\begin{equation}
	\begin{split}
	\bra{D_0}\widetilde q^{\frac12(L_0+\bar L_0-D/12)}\ket{D_0}
	&=2^{-D/2}\frac{\Theta_L(\widetilde\tau/2)}{\eta(\widetilde\tau)^D}\\
	&=\frac{\Theta_L(2\tau)}{\eta(\tau)^D}
	=\frac{\sum_{w\in L}q^{w^2}}{\eta(\tau)^D}.
	\end{split}
\end{equation}
Here the second equality follows from Poisson resummation and $L^*=L$, together with the modular transformation of $\eta$. The open-channel states have weights $w^2+N$, with $N\in\mathbb Z_{\ge0}$. Since $L$ is positive definite, weight zero requires $w=0$ and $N=0$. The corresponding oscillator vacuum occurs with multiplicity one in the open-channel partition function. This verifies the normalization for a single elementary Dirichlet brane and shows that the boundary condition is simple.

The lift~\eqref{eq:GonV} satisfies $D_k\ket{m,0}=\ket{r(k)m,0}$, so it preserves the momentum sum without additional phases. Orthogonality of $r(k)$ also preserves $\sum_i a_{-n}^i\bar a_{-n}^i$. Consequently,
\begin{equation}
	D_k\ket{D_0}=\ket{D_0}
	\qquad\text{for every }k\in K.
\end{equation}
For an invertible symmetry, invariance of a simple boundary means that each symmetry line can end topologically on it, with a one-dimensional space of endpoints. Choosing these endpoints gives comparison phases $\beta(k,\ell)$ between composing the endpoints for $k,\ell$ and using the endpoint for $k\ell$. Consistency of composing three endpoints requires $\delta\beta=\alpha|_K$, so the restricted anomaly is a coboundary. This is the finite-group boundary criterion of~\cite[Secs.~2.2--2.3 and 3.1]{ChoiRayhaunSanghaviShao2023}, whose assumption of a unique bulk vacuum also holds for $\mathcal N(L)$. We therefore obtain
\begin{equation}
	\operatorname{res}^{G}_{K}[\alpha]=0.
\end{equation}

\paragraph{Restriction to \texorpdfstring{$\langle t\rangle$}{the subgroup generated by t}}
We then consider $\operatorname{res}^{G}_{\langle t\rangle}[\alpha]$, for which it is more convenient to check the level-matching criterion since $\mathbb Z_2$ is cyclic. Since $H^3(\mathbb Z_2, U(1)) = \mathbb Z_2$, the restriction $\operatorname{res}^{G}_{\langle t\rangle}[\alpha]$ can take values in $\{0,1\}$, and the spins of states in the unprojected $\mathcal T$-twisted sector obey the selection rule~\cite{LinShao2021}
\begin{equation}
	2s = 2(h - \overline h) \in \begin{cases}
		\mathbb Z, & \operatorname{res}^{G}_{\langle t\rangle}[\alpha] = 0 \\
		\mathbb Z + \frac{1}{2}, & \operatorname{res}^{G}_{\langle t\rangle}[\alpha] = 1
	\end{cases}.
	\label{eq:LevelMatching_Narain}
\end{equation}
Thus $\operatorname{res}^{G}_{\langle t\rangle}[\alpha]$ vanishes precisely when the $\mathcal T$-twisted spectrum satisfies the level-matching condition $2(h-\bar h)\in\mathbb Z$ before projection~\cite[Eq.~(2.6)]{LinShao2021}. To compute this spectrum, given the convention $Z_{\text{spatial twist},\,\text{temporal twist}}$, we start from the trace over the untwisted Hilbert space $\mathcal H$ with $\Theta_L$ as above,
\begin{equation}
	Z_{1,\mathcal T} = \operatorname{Tr}_{\mathcal H} \left( \mathcal T\,q^{L_0-D/24}\bar q^{\bar L_0-D/24} \right)
	=\frac{\Theta_L(2\tau)}{\eta(\tau)^D} \overline{\left(\frac{\eta(\tau)}{\eta(2\tau)}\right)^D}.
\end{equation}
Indeed, since $\mathcal T\ket{m,w} = (-1)^{(m,w)}\ket{w,m}$, only the charge sectors with $m=w=\lambda$ contribute to the trace. Each of them carries the phase $(-1)^{(\lambda,\lambda)}=1$ by the evenness of $L$ and has charge contribution $(h,\bar h)=(\lambda^2,0)$, while by~\eqref{eq:GonO} every right-moving creation operator contributes a sign. A modular $S$ transformation $Z_{\mathcal T,1}(\tau,\bar\tau) = Z_{1,\mathcal T}(-1/\tau,-1/\bar\tau)$ turns the insertion in the trace into a spatial twist~\cite{DVVV1989}. Using $\Theta_L(-2/\tau) = (-i\tau/2)^{D/2}\Theta_L(\tau/2)$ and $\eta(-1/\tau) = \sqrt{-i\tau}\eta(\tau)$, we obtain
\begin{equation}
	Z_{\mathcal T,1} =\operatorname{Tr}_{\mathcal H_{\mathcal T}} \left(q^{L_0-D/24}\bar q^{\bar L_0-D/24}\right)
	=\frac{\Theta_L(\tau/2)}{\eta(\tau)^D} \overline{\left(\frac{\eta(\tau)}{\eta(\tau/2)}\right)^D}.
	\label{eq:ZT1_Narain}
\end{equation}
Since
\begin{equation}
	\begin{aligned}
    \frac{\Theta_L(\tau/2)}{\eta(\tau)^D} &= \sum_{\lambda\in L} q^{\lambda^2/4-D/24} \prod_{n\ge1}(1-q^n)^{-D},\\
    \left(\frac{\eta(\tau)}{\eta(\tau/2)}\right)^D &= q^{D/48} \prod_{n\ge0}(1-q^{n+1/2})^{-D},
    \end{aligned}
\end{equation}
the spectrum of $\mathcal H_{\mathcal T}$ is~\eqref{eq:T-spectrum_Narain}, with oscillator levels
\begin{equation}
    N_L=\sum_{i=1}^{D}\sum_{n\ge1}n\,n^L_{i,n}, \qquad N_R^{\mathrm{tw}} =\sum_{i=1}^{D}\sum_{n\ge0} \left(n+\frac12\right)n^R_{i,n+1/2}.
\end{equation}
Therefore, the spins satisfy
\begin{equation}
    2s=\frac{\lambda^2}{2}-\frac D8+2N_L-2N_R^{\mathrm{tw}}\in\mathbb Z
\end{equation}
since $L$ is even, $D\in8\mathbb Z$, $N_L\in\mathbb Z_{\ge0}$ and $N_R^{\mathrm{tw}}\in\frac12\mathbb Z_{\ge0}$. This agrees with the level-matching condition for asymmetric orbifolds by involutions of a Narain lattice~\cite[Eq.~(6.8)]{HarveyMoore2018}, specialized to no twisted left-moving bosons, $D$ twisted right-moving bosons, vanishing shift, and the dual $L/\sqrt2$ of the invariant lattice $\sqrt2L$ of $t$. Thus, by the level-matching condition~\eqref{eq:LevelMatching_Narain} we have $\operatorname{res}^{G}_{\langle t\rangle}[\alpha] = 0$.

We also note that the leading term of~\eqref{eq:ZT1_Narain} is $q^{-D/24}\bar q^{D/48}$ with coefficient one, coming from $\lambda=0$ and the oscillator vacua. Thus $\mathcal H_{\mathcal T}$ has a unique ground state, of weights $(0,D/16)$.

\paragraph{Mixed summand}
It remains to show that $[\alpha]$ pairs trivially with the mixed summand $H_2(K,\mathbb Z)\otimes\mathbb Z_2$ in~\eqref{eq:Kunneth_Narain}. To test this contribution, we choose a normalized cocycle $\alpha$ representing the anomaly. Since $t$ is central in $G$, every $k\in K$ preserves the $t$-twisted sector. Indeed, in general $k$ maps the sector twisted by $t$ to the sector twisted by $ktk^{-1}$, which here is again $\mathcal H_{\mathcal T}$. We choose an operator $U_k$ implementing the $k$-action on $\mathcal H_{\mathcal T}$. In the presence of the anomaly represented by $\alpha$, these operators form a projective representation of $K$ whose multiplier is the transgression of $\alpha$ along $t$~\cite{Willerton2008}
\begin{equation}
    U_kU_l=\beta_t(k,l)U_{kl},\qquad
    \beta_t(k,l)=
    \frac{\alpha(t,k,l)\alpha(k,l,t)}{\alpha(k,t,l)}.
    \label{eq:multiplier_Narain}
\end{equation}
We note that this transgression is the only one needed here. Indeed, the restrictions of the anomaly to the pure $H_3(K,\mathbb Z)$ and $H_3(\mathbb Z_2,\mathbb Z)$ summands of~\eqref{eq:Kunneth_Narain} have already been shown to vanish, while $[t]$ generates $H_1(\mathbb Z_2,\mathbb Z)$ in~\eqref{eq:Kunneth_Narain}. Slanting $[\alpha]$ with $[t]$ therefore detects the entire remaining summand $H_2(K,\mathbb Z)\otimes H_1(\mathbb Z_2,\mathbb Z)$. Since $U_k$ commutes with $L_0$ and $\bar L_0$, the one-dimensional ground-state space of $\mathcal H_{\mathcal T}$ found above is preserved by $K$. Therefore $U_k$ acts on this line by a scalar $u(k)$, and restricting~\eqref{eq:multiplier_Narain} to the line gives
\begin{equation}
    \beta_t(k,l)=\frac{u(k)u(l)}{u(kl)},\qquad
    [\beta_t]=0\in H^2(K,U(1))
\end{equation}
where $[\delta u] = [\beta_t] = [\iota_t\alpha]$. Thus for every $x\in H_2(K,\mathbb Z)$ we have
\begin{equation}
    \langle[\alpha],x\times[t]\rangle
    =\langle[\beta_t],x\rangle=1
    \label{eq:mixed-zero_Narain}
\end{equation}
for $\langle\cdot,\cdot\rangle\in U(1)$. Consequently the anomaly $[\alpha]$ vanishes on the mixed summand.

Combining the three steps, $[\alpha]$ pairs trivially with every summand in~\eqref{eq:Kunneth_Narain}. Applying the universal coefficient theorem leads to $[\alpha]=0$ in $H^3(G,U(1))$, which completes the proof of Theorem~\ref{thm:Narain-anomaly}.

\end{appendices}

\backmatter

\bmhead{Acknowledgements}
JT is supported by National Natural Science Foundation of China under Grant No. 12405085 and by the Natural Science Foundation of Shanghai (Grant No. 24ZR1419300). JT would like to thank Ying Zhang for her love and support.

\section*{Declarations}

\bmhead{Conflict of interest} The authors declare no competing interests.

\bmhead{Data availability} The authors declare that the data supporting the findings of this study are available within the paper.

\bibliography{apssamp}

\end{document}